\documentclass[journal]{IEEEtran}
\usepackage{ucs}
\usepackage[utf8x]{inputenc}
\usepackage[cmex10]{amsmath}
\usepackage{cite, amsfonts, amssymb, amsthm, bm, bbm, graphicx, relsize, multirow, booktabs, tikz,subfigure,soul}
\usepackage[american]{babel}
\usepackage[T1]{fontenc}
\usepackage{algorithmic, algorithm}
\usepackage{algorithm}
\usepackage[multiple]{footmisc}
\usepackage{glossaries}

\newcommand{\ds}{\displaystyle}

\newtheorem{proposition}{Proposition}
\newtheorem{lemma}{Lemma}

\theoremstyle{definition}

\theoremstyle{remark}

\makeglossaries

\newacronym{see}{SEE}{secrecy energy efficiency}
\newacronym{miso}{MISO}{multiple input single output}
\newacronym{miso-se}{MISO-SE}{multiple input single output single-antenna eavesdropper}
\newacronym{lmmse}{LMMSE}{linear minimum mean square error}
\newacronym{d2d}{D2D}{device-to-device}
\newacronym{p2p}{P2P}{point-to-point}
\newacronym{mac}{MAC}{multiple-access channel}
\newacronym{bc}{BC}{broadcast channel}
\newacronym{ic}{IC}{interference channel}
\newacronym{imac}{IMAC}{interference multiple access channel}
\newacronym{ibc}{IBC}{interference broadcast channel}
\newacronym{mimo}{MIMO}{multiple-input multiple-output}
\newacronym{mimo-me}{MIMO-ME}{multiple input multiple output multiple-antenna eavesdropper}
\newacronym{siso}{SISO}{single-input single-output}
\newacronym{sc}{SC}{single-carrier}
\newacronym{mc}{MC}{multi-carrier}
\newacronym{ofdma}{OFDMA}{orthogonal frequency division multiple access}
\newacronym{af}{AF}{amplify-and-forward}
\newacronym{df}{DF}{decode-and-forward}
\newacronym{cf}{CF}{compress-and-forward}
\newacronym{mwrc}{MWRC}{multi-way relay channel}
\newacronym{pde}{PDE}{partial data exchange}
\newacronym{fde}{FDE}{full data exchange}
\newacronym{iid}{i.i.d.\@}{independent and identically distributed}
\newacronym{awgn}{AWGN}{additive white Gaussian noise}
\newacronym{awg}{AWG}{additive white Gaussian}
\newacronym{sic}{SIC}{successive interference cancellation}
\newacronym{dpc}{DPC}{dirty paper coding}
\newacronym{snr}{SNR}{signal-to-noise ratio}
\newacronym{sinr}{SINR}{signal to interference plus noise ratio}
\newacronym{ber}{BER}{bit error rate}
\newacronym{zf}{ZF}{zero-forcing}
\newacronym{mmse}{MMSE}{minimum mean square error}
\newacronym{sud}{SUD}{single user decoding}
\newacronym{dof}{DoF}{degrees of freedom}
\newacronym{gdof}{GDoF}{generalized degrees of freedom}
\newacronym{nnc}{NNC}{noisy network coding}
\newacronym{dmn}{DMN}{discrete memoryless network}
\newacronym{csi}{CSI}{channel state information}
\newacronym{ee}{EE}{energy efficiency}
\newacronym{ian}{IAN}{treating interference as noise}
\newacronym{snd}{SND}{simultaneous non-unique decoding}
\newacronym{brd}{BRD}{best response dynamics}
\newacronym{br}{BR}{best response}
\newacronym{ne}{NE}{Nash equilibrium}
\newacronym{lhs}{LHS}{left-hand side}
\newacronym{rhs}{RHS}{right-hand side}
\newacronym{gee}{GEE}{global energy efficiency}
\newacronym{wsee}{WSEE}{weighted sum energy efficiency}
\newacronym{wpee}{WPEE}{weighted product energy efficiency}
\newacronym{wmee}{WMEE}{weighted minimum energy efficiency}
\newacronym{kkt}{KKT}{Karush Kuhn Tucker}
\newacronym{pc}{PC}{pseudo-concave}
\newacronym{qc}{QC}{quasi-concave}
\newacronym{ql}{QL}{quasi-linear}
\newacronym{pl}{PL}{pseudo-linear}
\newacronym{spc}{SPC}{strictly pseudo-concave}
\newacronym{sqc}{SQC}{strictly quasi-concave}
\newacronym{lfp}{LFP}{linear fractional problem}
\newacronym{clfp}{CLFP}{concave-linear fractional problem}
\newacronym{ccfp}{CCFP}{concave-convex fractional problem}
\newacronym{mmfp}{MMFP}{max-min fractional problem}
\newacronym{sorp}{SoRP}{sum-of-ratios problem}
\newacronym{porp}{PoRP}{product-of-ratios problem}
\newacronym{qos}{QoS}{quality of service}
\newacronym{evd}{EVD}{eigenvalue decomposition}
\newacronym{svd}{SVD}{singular value decomposition}
\newacronym{skee}{SKEE}{Secret-key energy efficiency}
\newacronym{an}{AN}{artificial noise}

\begin{document}

\bstctlcite{IEEEexample:BSTcontrol}

This paper was submitted for publication on the IEEE Transactions on Green Communications and Networking on April 24, 2018 and was assigned reference number TGCN-TPS-18-0048.
It was finally accepted for publication on March 24, 2019.  

\bigskip

\copyright 2019 IEEE. Personal use of this material is permitted. Permission from IEEE must be obtained for all other uses, in any current or future media, including reprinting/republishing this material for advertising or promotional purposes, creating new collective works, for resale or redistribution to servers or lists, or reuse of any copyrighted  component of this work in other works.”

\title{Energy-Efficient Power Control in Cell-Free and User-Centric Massive MIMO at Millimeter Wave}
\author{Mario Alonzo, Stefano Buzzi, {\em Senior Member}, {\em IEEE},  Alessio Zappone {\em Senior Member,  IEEE}, and Ciro D'Elia
\thanks{This paper was partly presented at the 28th Annual IEEE International Symposium on Personal, Indoor and Mobile Radio Communications, Montreal (Canada), October 2017 and at the 2018 IEEE 5G World Forum, Santa Clara, CA, July 2018. The work of M. Alonzo, S. Buzzi, and C. D'Elia  has been supported by the Italian Ministry of Education and Research, under the program "Dipartimenti di Eccellenza 2018-2022". The work of A. Zappone was supported
by H2020 MSCA IF BESMART, Grant 749336.}
\thanks{M. Alonzo, S. Buzzi and C. D'Elia are with the Department of Electrical and Information Engineering, University of Cassino and Lazio Meridionale, I-03043 Cassino, Italy (mario.alonzo@unicas.it, buzzi@unicas.it, delia@unicas.it). S. Buzzi and C. D'Elia are also affiliated with CNIT, Parma, Italy. Alessio Zappone is with the LANEAS group of the L2S, CentraleSupelec, CNRS, UnivParisSud, Universit\'e Paris-Saclay, 91192 Gif-sur-Yvette, France. (alessio.zappone@l2s.centralesupelec.fr).}}
\maketitle
\thispagestyle{empty}

\begin{abstract}
In a cell-free massive MIMO architecture a very large number of distributed access points simultaneously and jointly serves a much smaller number of mobile stations; a variant of the cell-free technique is the user-centric approach, wherein each access point just serves a reduced set of mobile stations. This paper introduces and analyzes the cell-free and user-centric architectures at millimeter wave frequencies, considering a training-based channel estimation phase, and the downlink and uplink data transmission phases.  First of all, a multiuser clustered millimeter wave channel model is introduced in order to account for the correlation among the channels of nearby users; second, an uplink multiuser channel estimation scheme is described along with low-complexity hybrid analog/digital beamforming architectures. Third, the non-convex problem of power allocation for downlink global energy efficiency maximization is addressed.
Interestingly, in the proposed schemes no channel estimation is needed at the mobile stations, and the beamforming schemes used at the mobile stations are channel-independent and have a very simple structure. Numerical results show the benefits granted by the power control procedure, that the considered architectures are effective, and permit assessing the loss incurred by the use of the hybrid beamformers and by the channel estimation errors.  
\end{abstract}


\section{Introduction}
Future 5G wireless systems will heavily rely on the use of large-scale antenna arrays, a.k.a. massive MIMO, and of carrier frequencies above $10\,\text{GHz}$, the so called mmWave frequencies \cite{WhatWill5gbe}. Indeed, on one hand, the use of massive MIMO permits serving several users on the same time-frequency resource, while, on the other one, mmWave carrier frequencies will enable the use of much larger bandwidths, at least on short-distances (up to about one hundred meters). The combined use of massive MIMO systems along with mmWave frequencies is indeed one of the key technological enablers of the envisioned wireless Gbit/s experience \cite{swindlehurst2014millimeter}. For conventional sub-$6\,\text{GHz}$ frequencies, a new communications architecture, named CF massive MIMO, has been recently introduced in \cite{CFNgo1,CFNgo2}, in order to alleviate the cell-edge problem and thus increase the system performance of unlucky users that happen to be located very far from their serving AP. In the CF architecture, instead of having few base stations with massive antenna arrays, a very large number of simple APs randomly and densely deployed serve a much smaller number of MSs.
This approach has some similarities with the CoMP \cite{irmer2011coordinated,sawahashi2010} and network MIMO \cite{venkatesan2007}; these technologies exploit coordinate beamforming, scheduling, and  joint transmission using multiple distributed antennas and/or base stations, with the aim of mitigating the interference and achieving diversity gains. There are however some  key differences between these strategies and the CF massive MIMO approach discussed in this paper. In particular, we have that: (a) the CF massive MIMO strategy fully leverages the advantages and unique features of massive MIMO, such as the channel hardening effect and the use of the TDD protocol;and (b) the CF massive MIMO also makes a more limited use of the backhaul link since the channel coefficients are locally estimated at the APs and are mutually shared, and, moreover, beamformers are computed at the APs using locally available information.
The CF massive MIMO architecture can be thus thought as the scalable way of implementing distributed network MIMO deployments.  
 In the CF architecture described in \cite{CFNgo1,CFNgo2}, single-antenna APs and MSs are considered, all the APs serve all the MSs, all the APs are connected through a backhaul link to a CPU, but every AP performs channel estimation locally, and channel estimates are not sent to the CPU, but are locally exploited. In particular, for the downlink communication phase, the CPU sends to the APs the data symbols to be sent to all the MSs, while for the uplink communication phase, the APs use the backhaul to send the sufficient statistics for all the MSs to the CPU, which combines them and performs uplink data decoding.  
In \cite{buzzi2017cell} a UC variant of the CF approach is introduced, wherein each APs, instead of serving all the MSs in the considered area, just serves the ones that he receives best; the results of \cite{buzzi2017cell} show that the UC approach provides savings on the required backhaul capacity and, also, provides better data-rates to the vast majority of the users. 
\begin{table}[t]
\caption{List of acronyms}
\center
\begin{tabular}{|l|l|}
\hline
5G    & Fifth-Generation               \\ \hline
AP   & Access-Point                   \\ \hline
BCD-SD & Block Coordinate Descent for Subspace Decomposition \\ \hline
CF   & Cell-Free                      \\ \hline
CoMP & Coordinated Multipoint         \\ \hline
CPU   & Central Processing Unit                   \\ \hline
CSI   & Channel State Information                  \\ \hline
FD & Fully-Digital \\ \hline
HY & Hybrid \\ \hline
LMMSE & Linear Minimum Mean Square Error \\ \hline 
LTE  & Long-Term Evolution            \\ \hline
MIMO & Multiple-Input-Multiple-Output \\ \hline
mmWave & Millimeter Wave \\ \hline 
MS   & Mobile Station                 \\ \hline
TDD & Time Division Duplex \\ \hline 
ULA & Uniform Linear Array \\ \hline
ZF & Zero-Forcing \\ \hline
\end{tabular}
\end{table}
The paper \cite{Total_EE} analyzes a CF massive MIMO system from the point of view of its energy consumption, taking into account the power consumed by the backhaul links, the number of active APs and the number of antennas at each AP. The paper also embraces the UC philosophy by considering AP-MS selection schemes aimed at maximizing the system energy efficiency. While previous papers deal with the case in which the APs and the MSs are both equipped with one antenna (with the exception of \cite{Total_EE} which considers multiple antennas at the APs), in \cite{WSA2017_buzzi_dandrea_CF} the CF and UC  architectures are generalized to the case in which both the APs and the MSs are equipped with multiple antennas. This generalization is not trivial and indeed the system proposed in \cite{WSA2017_buzzi_dandrea_CF}, 
despite the use of multiple antennas at the MSs and the use of a multiplexing order larger than one, develops a beamforming scheme at the APs that does not require channel estimation at the MSs, which  adopt a channel-independent beamformer.
While \cite{WSA2017_buzzi_dandrea_CF} considers a simple uniform power allocation, in \cite{Buzzi_Zappone_PIMRC2018,buzzi2018user} power control procedures aimed at rate maximization are proposed.
The paper \cite{cell-free_downlinkpilots} investigates the use of pilot signals on the downlink, so as to enable channel estimation at the MSs; even though such an approach provides some performance improvement, it contradicts the TDD-based "Massive MIMO philosophy" wherein no channel estimation is required at the MSs. A compute-and-forward approach to CF massive MIMO was then proposed in \cite{huang2017compute} in order to reduce the load on the backhaul links. 

All of the above cited papers consider the case in which a conventional sub-6 GHz carrier frequency is used. On the other hand, as already stated, mmWave frequencies will play a big role in future wireless cellular systems due to availability of wide amounts of unused bandwidths \cite{6515173}. The design of a wireless cellular system operating at mmWave poses different and new challenges with respect to conventional sub-6 GHz frequencies, due to different propagation mechanisms \cite{6515173}, to the need of using large antenna arrays at both side of the links to counteract the increased 
path-loss \cite{buzzi2018energy_TGCN}, and to the presence of hardware constraints that prevent the realization of FD precoding and postcoding beamforming structures \cite{alkhateeb2014mimo}. Despite these tough  challenges, the attractiveness of the mmWave frequencies for cellular communications has led to intense research efforts in the last few years. 
Due to the difficulty of realizing FD beamforming architectures at mmWave with antenna arrays of large size, the CF massive MIMO architecture at these high frequencies appears particularly attractive, mainly for two reasons; first of all, it substitutes large co-located antennas with several APs 
with antenna arrays of smaller dimension, and, thus, with less hardware complexity; second, due to the limited range of mmWave communications, the distributed AP dense deployment alleviates the cell-edge problem and creates path-diversity, which is precious since signal blockages may frequently happen at these frequencies. 

Energy efficiency is another key topic that should be considered when designing modern wireless communication systems. Indeed, increasing the  bit-per-Joule energy efficiency is regarded as a key requirement of future 5G networks \cite{5GNGMN}. A recent survey on the most promising energy-efficient techniques for 5G has recently  appeared in \cite{GEJSAC16}; in that paper,  radio resource allocation is identified, among other techniques, as a relevant approach to improve the energy efficiency of future wireless networks. Recent contributions on energy-efficient radio resource allocation for 5G are \cite{yun2014energy,ZapNow15,ZapTSP15,pizzo2017optimal}.

Motivated by the above discussion, this paper considers details a CF and UC massive MIMO wireless system operating at mmWave and focuses on resource allocation strategies maximizing the global energy efficiency. To the best of authors'  knowledge, this is the first paper that considers the CF and UC massive-MIMO deployments at mmWave; preliminary results on this issue have been reported by the authors in the conference papers \cite{alonzo2017cell} and \cite{alonzo5GWF},  wherein it has also been shown that the UC approach generally
outperforms the CF approach.

The contribution of the paper can be summarized as follows. First of all, we introduce a multiuser mmWave channel model that permits taking into account channel correlation for close users. Building upon the clustered channel model \cite{buzzidandreachannel_model} widely used at mmWave frequencies, we extend this model to take into account the fact that if several APs and MSs are in the same area, their channels must be built using the same set of scatterers; adopting this model, users that are very close will receive beams with very close direction of arrival and so channel correlation for nearby users is intrinsically taken into account. Then, we study the UC and CF approaches at mmWave frequencies; we assume that both the APs and MSs are equipped with multiple antennas, use HY analog-digital partial ZF beamforming at the APs, while a very simple 0-1 beamforming architecture, independent of the channel estimate, is used at the MSs. 
We present simulation results for scenarios that can be representative of a lightly-loaded system, with $M=80$ APs and $K=6$ MSs transmitting on the same time-frequency slot,  and of a highly-loaded system, with $M=80$ APs and $K=16$ MSs.
Results show that over a bandwidth of $200\,\textrm{MHz}$ and with a maximum transmit power of $0.1\,\textrm{W}$ at each AP, taking into account channel estimation errors and using low-complexity beamforming structures, the downlink average rate per user is $1.5\,\textrm{Gbit/s}$ in the lightly-loaded scenario and $400\,\textrm{Mbit/s}$ in the heavily-loaded situation. Likewise,  in the lightly-loaded scenario, the average uplink rate-per-user is about $1\,\textrm{Gbit/s}$. 

This paper is organized as follows. Next section is devoted to the discussion of the system and to the description of the processing (it involves uplink training, downlink data transmission and uplink data transmission); Section \ref{Section:GEE} concerns global energy efficiency maximization; Section \ref{Section:rate_maximization} concerns the sum-rate optimization under (theoretical) perfect ZF beamforming; Section \ref{Section:Numerical_results} contains details about the used multiuser channel model and the discussion of the numerical results. Finally, concluding remarks are given in Section \ref{Section:Conclusion}.

\medskip
\textit{Notation. }We use lower case boldface characters to denote column
vectors and capital boldface characters to denote matrices.
We denote $N \times N$ identity matrix with $\textbf{I}_{N}$; $(\cdot)^{T}$, $(\cdot)^{H}$ and $(\cdot)^{*}$ denote the transpose, conjugate transpose and complex conjugate of a matrix. Then we denote with $\text{tr}(\textbf{A})$, $\parallel \textbf{A} \parallel$, $\mid \textbf{A} \mid$ the trace, the norm and the determinant of a matrix $\textbf{A}$. $\textbf{x}\sim\mathcal{CN}(\mu,\sigma^2)$ indicates a Complex Gaussian random variable with mean $\mu$ and variance $\sigma^{2}$. $\otimes$ indicate the Kronecker product between two matrices. We denote with $\text{card}(\cdot)$ a set's cardinality. A list of used acronyms is reported in Table I.

\section{System model and communication protocol}
\label{section:System_Model}
\begin{figure}[h]
\centering
\includegraphics[scale=0.6]{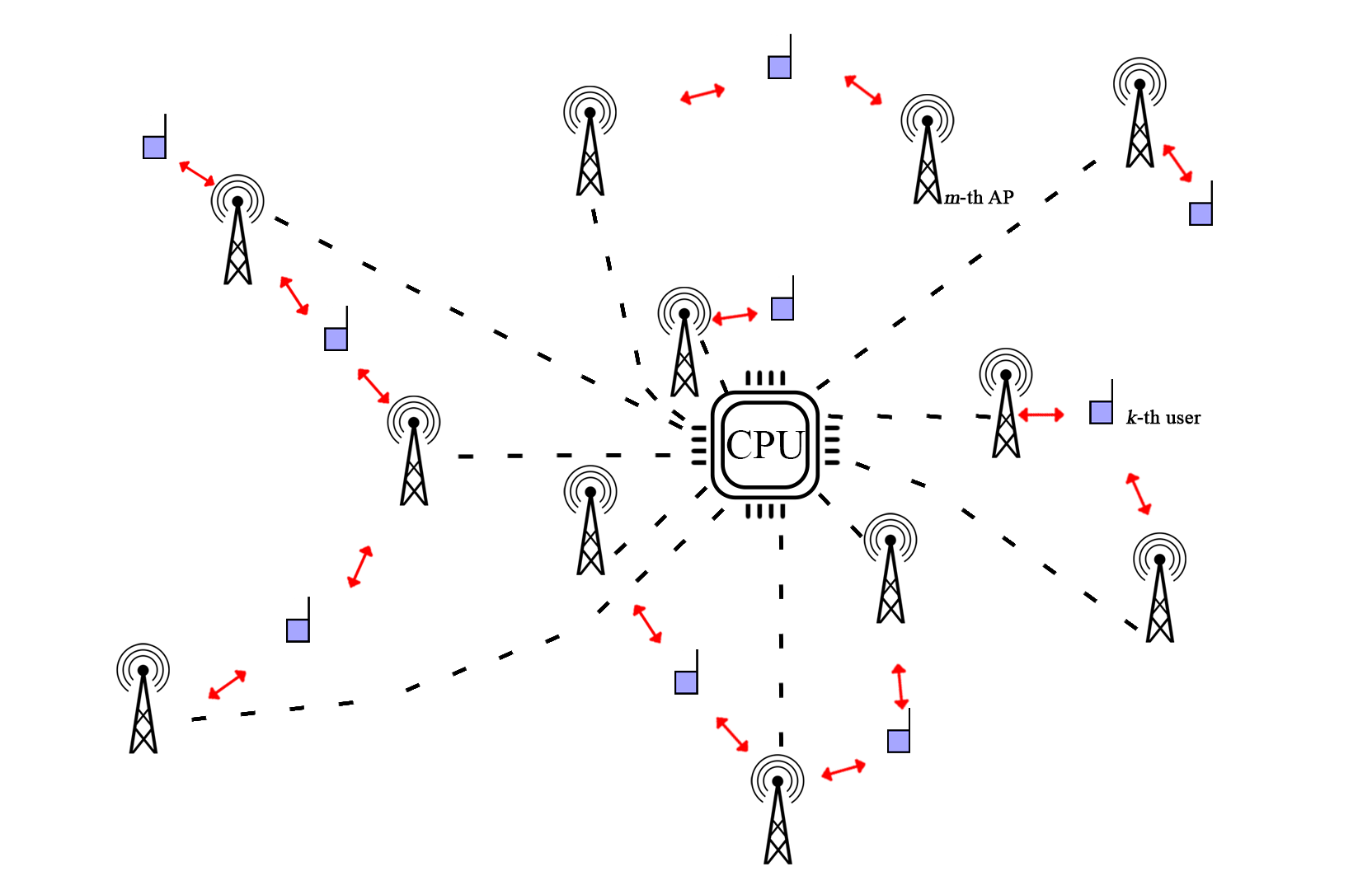}
\caption{A CF massive MIMO network deployment.}
\label{fig:cf}
\end{figure}

We consider an area where $K$ MSs and  $M$ APs are randomly located.
The APs are connected by means of a backhaul network to a CPU wherein data-decoding is performed (see fig. \ref{fig:cf}). Communications take place on the same frequency band; downlink and uplink are separated through TDD\footnote{In TDD, using calibrated hardware, the uplink channel is the reciprocal of downlink channel.}. The communication protocol is made of three different phases: uplink training, downlink data
transmission and uplink data transmission. During the uplink training phase, the MSs send pilot sequences to the APs and each AP estimates the channels; during the second phase the APs use the channel estimates to perform pre-coding and transmit the data symbols; finally, in the third phase the MSs send uplink data symbols to the APs. 
While in the CF approach all the APs simultaneously serve all the MSs (a fully-cooperative scenario), in the UC approach each  AP serves a pre-determined number of MSs, say $N$, and in particular the ones that it receives best.
We assume that each AP (MS) is equipped with an antenna array with $N_{AP}$ ($N_{MS}$) elements. 
The ($N_{AP}\times N_{MS}$)-dimensional matrix $\textbf{H}_{k,m}$ denotes the channel matrix between the $k$-th user and the $m$-th AP. Details about the channel model will be reported in Section \ref{Section:Numerical_results}

Additionally, we assume that each MS employs a very simple 0-1 beamforming structure; in particular, denoting by $P$ the multiplexing order, namely the number of parallel streams sent to a given receiver, 
 the $(N_{MS}\times P)$-dimensional beamformer used at the $k$-th MS receiver is denoted by $\textbf{L}_k$ and is defined as
 $\textbf{L}_k=\textbf{I}_P\otimes \textbf{1}_{N_{MS}/P}$, denoting with $\textbf{1}_{N_{MS}/P}$ an all-1 vector of length $N_{MS}/P$. Otherwise stated, we assume that the MS receive antennas are divided in $P$ disjoint groups of $N_{MS}/P$ elements, and the received data collected at the antennas of each group are simply summed together. It is APs' task, based on the uplink channel estimates and exploiting the TDD channel reciprocity, to ensure that the summed samples are, at least approximately, aligned in phase. 
Similarly when considering uplink transmission, the antennas in each group send the same signal with the same phase. 
 We describe now the three phases of the communication protocol.

\subsection{Uplink training}
During the uplink training the MSs transmit pilot sequences in order to enable channel estimation at the APs. Let 
$\tau_c$ be the length  of the channel coherence time and $\tau_p$ be the length of uplink training phase, both in discrete time samples.  Of course we must have $\tau_p < \tau_c$. We define by $\boldsymbol{\Phi}_k \in \mathcal{C}^{P\times \tau_p}$ the matrix containing on its rows the pilot sequences sent by the $k$-th MS. We assume that 
$\boldsymbol{\Phi}_k\boldsymbol{\Phi}_k^H =\textbf{I}_P$,  i.e. the rows of $\boldsymbol{\Phi}_k$ are orthogonal, but no orthogonality is required for the pilot sequences assigned to other MSs\footnote{Of course, when $KP\leq \tau_p$ it would be possible to assign to all the MSs mutually orthogonal pilot sequences. In general, the pilot sequences  contain symbols from the same constellation as the data symbols, e.g. QAM symbols or similar. In this paper, however, for the sake of simplicity, we assume that the pilot sequences are binary random sequences, and we just require that each matrix $\boldsymbol{\Phi}_k$ has orthogonal rows.}. Obviously, using orthogonal pilots \textit{tout court} would lead to a system immune to pilot contamination, but this would put a limit on the maximum value of the product $KP$ that could be accommodated in the channel coherence time. 
  The received signal at the $m$-th AP in the $\tau_p$ signaling intervals devoted to uplink training can be cast in the following  $N_{AP}\times \tau_p$-dimensional matrix $\textbf{Y}_m$: 
  \begin{equation}
  \label{eq:Y_m_training}
  \textbf{Y}_{m}=\sum\limits_{k=1}^{K}\sqrt{p_k}\textbf{S}_{k,m}\boldsymbol{\Phi}_k+\textbf{W}_m,
  \end{equation}
  where $\textbf{S}_{k,m}=\textbf{H}_{k,m}\textbf{L}_{k}$, $\textbf{W}_m$ is the matrix of thermal noise samples, whose entries are assumed to be i.i.d. $\mathcal{CN}(0,\sigma_w^2)$ RVs. 
  In the following we briefly outline the structure of the LMMSE channel estimator.
  Defining $\textbf{y}_{m}=\text{vec}(\textbf{Y}_{m})$, $\textbf{w}_{m}=\text{vec}(\textbf{W}_{m})$, $\textbf{s}_{k,m}=\text{vec}(\textbf{S}_{k,m})$, we obtain the vectorized model: 
  \begin{equation}
  \textbf{y}_{m}=\sum\limits_{k=1}^{K}\sqrt{p_{k}}\textbf{A}_{k}\textbf{s}_{k,m}+\textbf{w}_{m}
  \label{eq:vec_Y_m}
  \end{equation}
  with $\textbf{A}_{k}=\Phi_{k}^{T} \otimes \textbf{I}$. Since we need to estimate a vector, we process $\textbf{y}_{m}$ by a matrix $\textbf{V}_{k,m}^{H}$, i.e. $\hat{\textbf{s}}_{k,m}=\textbf{\textbf{}V}_{k,m}^{H}\textbf{y}_{m}$.  
  Then, the MSE is:
  \begin{equation}
  \begin{split}
  &\mathbb{E}[\parallel \textbf{V}_{k,m}^{H}\textbf{y}_{m} - \textbf{s}_{k,m} \parallel^{2}]=\text{tr}(\textbf{V}_{k,m}^{H}\mathbb{E}[\textbf{y}_{m}\textbf{y}_{m}^{H}]\textbf{V}_{k,m})+\\&+\mathbb{E}[\parallel \textbf{s}_{k,m}\parallel ^{2}]-\mathbb{E}[2\Re\{\text{tr}(\textbf{s}_{k,m}^{H}\textbf{V}_{k,m}^{H}\textbf{y}_{m})\}]=\\&=\text{tr}\bigg(\textbf{V}_{k,m}^{H}\bigg(\sum\limits_{l=1}^{K}p_{k}\textbf{A}_{l}\textbf{A}_{l}^{H}+\sigma^{2}\textbf{I}\bigg)\textbf{V}_{k,m}\bigg)+\\&+\mathbb{E}[\parallel \textbf{s}_{k,m}\parallel ^{2}]-\sqrt{p_{k}}\text{tr}(\textbf{V}_{k,m}^{H}\textbf{A}_{k}+\textbf{V}_{k,m}^{T}\textbf{A}_{k}^{*})
  \end{split}
  \label{eq:MSe}
  \end{equation}
Recalling that   (\cite[Ch.~4]{hjorungnes2011complex})   
 $ \nabla_{\textbf{Z}^{*}} \text{tr}(\textbf{Z}^{T}\textbf{M})=\textbf{0}$, $\nabla_{\textbf{Z}^{*}} \text{tr}(\textbf{Z}^{H}\textbf{M})=\textbf{M}$, and $\nabla_{\textbf{Z}^{*}} \text{tr}(\textbf{Z}^{H}\textbf{M}\textbf{Z})=\textbf{MZ}$, setting 
the gradient of the MSE with respect to the complex matrix $\textbf{V}_{k,m}^{*}$ equal to zero and solving for   $\textbf{V}_{k,m}$, we find the LMMSE estimator:
  \begin{equation}
  \textbf{V}_{k,m}^{LMMSE}=\sqrt{p_{k}}\big(\sum\limits_{l=1}^{K}p_{l}\textbf{A}_{l}\textbf{A}_{l}^{H}+\sigma^{2}\textbf{I}\big)^{-1}\textbf{A}_{k} \; .
  \label{eq:C_LMMSE}
  \end{equation}

\begin{algorithm}[t]
	\caption{Block Coordinate Descent for Subspace Decomposition Algorithm for Hybrid Beamforming; the algorithm input is $\mathbf{Q}_m^{opt}, N_{AP},  N_{AP}^{RF}, K, P$.}
	\label{algorithm:BCD_SD}
	\begin{algorithmic} [1]
		\STATE Initialize $I_{max}$ and set i=0;
		\STATE Set arbitrary $\textbf{Q}_m^{RF,0}$
		\REPEAT
		\STATE Update $\textbf{Q}_m^{BB}=(\textbf{Q}_m^{RF,i \, H}\textbf{Q}_m^{RF,i})^{-1}\textbf{Q}_m^{RF,i\, H}\textbf{Q}_m^{opt}$
		\STATE Set $\phi_{i}\!=\!{\rm angle}\!\!\left[\textbf{Q}_m^{opt}\textbf{Q}_m^{BB,i+1 \,H}(\textbf{Q}_m^{BB,i+1}\textbf{Q}_m^{BB,i+1 \,H})^{-1}\!\!\right]$
		\STATE Update $\textbf{Q}_m^{RF,i}=\frac{1}{\sqrt{N_{AP}}}e^{j \phi_{i}}$
		\STATE $i=i+1$; 
		\UNTIL{convergence or $i=I_{max}$}
	\end{algorithmic}
\end{algorithm}

\subsection{Downlink data transmission}
After the first phase, the generic $m$-th AP has an estimate of the quantities $\textbf{S}_{k,m}$, for all $k=1, \ldots, K$. 
In order to transmit data on the downlink, a ZF precoder is considered.
The precoding matrix $\mathbf{Q}_{k,m}$ is designed as follows.
 First of all, the matrix 
$\widehat{\mathcal{G}}=[\widehat{\textbf{S}}_{1,1}\dots \widehat{\textbf{S}}_{K,M}]$ is built;  then, we have
\begin{equation}
\textbf{Q}_{k,m}=(\widehat{\mathcal{G}}\widehat{\mathcal{G}}^H)^{-1}
\widehat{\textbf{S}}_{k,m}.
\end{equation}
Finally, each precoding matrix is normalized as follows: 
\begin{equation}
\textbf{Q}_{k,m}=\dfrac{\textbf{Q}_{k,m}}{\sqrt{\mbox{tr}(\textbf{Q}_{k,m}\textbf{Q}_{k,m}^{ H})}} \; ,
\end{equation}
$ \forall k=1,\dots,K,  \forall m=1,\dots,M$.

The previously described beamforming matrix  requires a FD implementation, which presumes the use of a number of RF chains equal to the number of transmit antennas. It is well-known that at mmWave frequencies hardware complexity constraints usually prevent the use of FD architectures, and thus HY beamforming structures have been proposed, where a number of RF chains $N_{AP}^{RF}< N_{AP}$ is used.  
In this paper we exploit the \emph{Block Coordinate Descent algorithm} \cite{Hybrid_BCDSD} in order to decompose our beamformer in the cascade of a FD (baseband) one, represented by a $(N_{AP}^{RF} \times P)$-dimensional matrix and of an analog one, represented by a $(N_{AP}\times N_{AP}^{RF})$-dimensional matrix whose entries have all constant norm. While the  baseband beamformer is MS-dependent, the analog beamformer  at the AP is unique and is used for transmitting to all the users. 
In particular, at the generic $m$-th AP, the following matrix, of dimension $N_{AP}\times KP$, is 
formed\footnote{In the UC approach, to be detailed in the following, the matrix $\mathbf{Q}_{m}^{opt}$ is
formed using the beamformers relative to the MS actually served by the $m$-th AP. }
\begin{equation}
\mathbf{Q}_{m}^{opt}=\left[\mathbf{Q}_{1,m}, \ldots, \mathbf{Q}_{K,m}\right] \; ,
\label{eq:Qopt}
\end{equation}
and used as an input to the algorithm \ref{algorithm:BCD_SD} in order to obtain the $N_{AP}\times N_{AP}^{RF}$ matrix $\mathbf{Q}_m^{RF}$ and the $N_{AP}^{RF}\times KP$ matrix 
\[
\mathbf{Q}_m^{BB}=\left[\mathbf{Q}_{1,m}^{BB}, \ldots, \mathbf{Q}_{K,m}^{BB} \right]\; ,
\] 
containing the MS-specific baseband beamformers to be used at the $m$-th AP. At the generic $m$-th AP, we will thus have as many digital beamformers as the MSs to transmit to, and only one analog beamformer, that will be used to transmit jointly to all the users. Although we are here considering, for the sake of simplicity, only one single frequency, it should be noted that with a multicarrier modulation a single analog beamformer must be used for all the subcarriers, or, if complexity permits, for each properly defined subset of contiguous subcarriers \cite{Hybrid_BCDSD}.

\subsubsection{The CF approach} In this case, all the APs serve all the MSs, so the transmitted signal from the \emph{m}-th AP in the \emph{n}-th sample interval is: 
\begin{equation}
\textbf{s}_{m}^{CF}(n)=\sum\limits_{k=1}^{K}\sqrt{\eta_{m,k}}\textbf{Q}_{k,m} \textbf{x}_{k}^{DL}(n),
\label{eq:s_m}
\end{equation} 
where $\textbf{x}_{k}^{DL}(n)$ is the data symbol intended for the $k$-th MS, and  $\eta_{m,k}$ is a scalar coefficient taking into account the power used by the $m$-th AP to transmit towards the $k$.th MS.  
The $k-th$ MS receives the following ($N_{MS} \times 1$)-dimensional vector: 
\begin{equation}
\label{eq:r_k}
\begin{split}
\textbf{y}_{k}^{CF}(n)&= \sum\limits_{m=1}^{M}\textbf{H}_{k,m}^{H}\textbf{s}_{m}^{CF}(n) + \textbf{z}_{k}(n)=\\&=\sum\limits_{m=1}^{M}\sqrt{\eta_{m,k}}\textbf{H}_{k,m}^{H}\textbf{Q}_{k,m} \textbf{x}_{k}^{DL}(n)+\\
&+\sum\limits_{l=1, l\neq{k}}^{K}\sum\limits_{m=1}^{M}\sqrt{\eta_{m,l}}\textbf{H}_{k,m}^{H}\textbf{Q}_{l,m} \textbf{x}_{l}^{DL}(n) + \textbf{z}_{k}(n),
\end{split}		
\end{equation}
where $\textbf{z}_{k}(n)$ is the additive thermal noise distributed as $\mathcal{CN}(0,\sigma_{z}^2)$.
A  soft estimate of the $k$-th MS data symbol is thus formed as 
\begin{equation}
\label{eq:x_k}
\widehat{\textbf{x}}_{k}^{DL,CF}(n)=\textbf{L}_{k}^{H}\textbf{y}_{k}^{CF}(n).
\end{equation}

\subsubsection{The UC approach} In this case  the APs are assumed to serve a pre-determined, fixed number of MSs, say $N$; in particular, we assume that the generic $m$-th AP serves the $N$ MSs whose channels have the largest Frobenious norms.  We denote by $\mathcal{K}(m)$ the set of MSs served by the $m$-th AP. Given the sets $\mathcal{K}(m)$, for all $m = 1,\dots,M$, we can define the set $\mathcal{M}(k)$ of the APs that communicate with the $k$-th user: 
\begin{equation}
\mathcal{M}(k)=\{m:k\in\mathcal{K}(m)\}
\label{set M_k}.
\end{equation}
In this case the transmitted signal from the $m$-th AP is written as: 
\begin{equation}
\textbf{s}_{m}^{UC}(n)=\sum\limits_{k\in\mathcal{K}(m)}\sqrt{\eta_{m,k}}\textbf{Q}_{k,m}\textbf{x}_{k}^{DL}(n).
\label{s_UC}
\end{equation}
The received signal at the $k$-th MS is expressed now as: 
\begin{equation}
\label{eq:r_k_Uc}
\begin{split}
\textbf{y}_{k}^{UC}(n)&= \sum\limits_{m=1}^{M}\textbf{H}_{k,m}^{H}\textbf{s}_{m}^{UC}(n) + \textbf{z}_{k}(n)=\\
&=\sum\limits_{m\in\mathcal{M}(k)}\sqrt{\eta_{m,k}}\textbf{H}_{k,m}^{H}\textbf{Q}_{k,m} \textbf{x}_{k}^{DL}(n) +\\
&+\sum\limits_{l=1, l\neq{k}}^{K}\sum\limits_{m\in\mathcal{M}(l)}\sqrt{\eta_{m,l}}\textbf{H}_{k,m}^{H}\textbf{Q}_{l,m} \textbf{x}_{l}^{DL}(n) + \textbf{z}_{k}(n),
\end{split}		
\end{equation}
As before the $N_{MS}$-dimensional vector $\textbf{z}_{k}(n)$ represents the thermal noise at the $k$-th MS, and it is modeled as i.i.d. $\mathcal{CN}(0,\sigma_{z}^{2})$. Then it is possible to obtain a soft estimate of the data symbol $\textbf{x}_{k}^{DL}(n)$ at $k$-th MS as:
\begin{equation}
\label{eq:x_hat_UC}
\hat{\textbf{x}}_{k}^{DL,UC}(n)=\textbf{L}_{k}^{H}\textbf{y}_{k}^{UC}(n).
\end{equation}

\medskip

Given the above equations, and assuming the use of Gaussian distributed long codewords, it is now possible to write down the achievable  rate for the \emph{k}-th user in the UC approach as follows: 
\begin{equation}
\mathcal{R}_{k}=\text{B}\hspace{0.05cm}\text{log}_{2}\hspace{0.05cm}\mid\textbf{I}+\textbf{R}_{k}^{-1}\textbf{A}_{k,k}\textbf{A}_{k,k}^{H}\mid \; , 
\label{eq:rate_k}
\end{equation}
where
\begin{equation}
\textbf{A}_{k,l}=\sum\limits_{m\in\mathcal{M}(k)}\sqrt{\eta_{m,l}}\textbf{L}_{k}^{H}\textbf{H}_{k,m}^{H}\textbf{Q}_{l,m} \; .
\label{A_kl}
\end{equation}
and 
\begin{equation}
\textbf{R}_{k}=\sum\limits_{l\neq k}\textbf{A}_{k,l}\textbf{A}_{k,l}^{H}+\sigma_{z}^{2}\textbf{L}_{k}^{H}\textbf{L}_{k}\, ,
\label{eq:cov_matrix}
\end{equation}
is the covariance matrix of the interference in the signal received at the $k$-th MS.
Now, some algebraic manipulations are needed in order to express the achievable rate $\mathcal{R}_{k}$ 
in \eqref{eq:rate_k} in a form that permits solving the optimization problems considered in Sections 
\ref{Section:GEE} and \ref{Section:rate_maximization}.
By letting:
\begin{equation}
\textbf{B}_{k,l,m}=\textbf{L}_{k}^{H}\textbf{H}_{k,m}^{H}\textbf{Q}_{l,m} \Longrightarrow 				      \textbf{A}_{k,l}=\sum\limits_{m\in\mathcal{M}(k)}\sqrt{\eta_{m,l}}\textbf{B}_{k,l,m}
\label{eq:B_klm}
\end{equation}
the covariance matrix \eqref{eq:cov_matrix}  can be rewritten as
\begin{equation}
	\begin{split}
\textbf{R}_{k}&=\sum\limits_{l\neq k}\sum\limits_{m\in\mathcal{M}(l)}\sum\limits_{m'\in\mathcal{M}(l)}\sqrt{\eta_{m,l}\eta_{m',l}}\textbf{B}_{k,l,m}\textbf{B}_{k,l,m'}^{H}+\\&+\sigma_{z}^{2}\textbf{L}_{k}^{H}\textbf{L}_{k}\; ,
\end{split}
\label{eq:cov_matrix1}
\end{equation}
thus implying that the rate for the \emph{k}-th MS can be  expressed as: 
\begin{equation}
\mathcal{R}_{k}=\text{B}\hspace{0.05cm}\text{log}_{2}\hspace{0.05cm}\begin{vmatrix}\textbf{I}+\textbf{R}_{k}^{-1}\sum\limits_{m,m'}\sqrt{\eta_{m,k}\eta_{m',k}}\textbf{B}_{k,k,m}\textbf{B}_{k,k,m'}^{H}\end{vmatrix}
\label{eq:rate_k1}
\end{equation}
Finally, exploiting the fact that the determinant of the product of two matrices factorizes into  the product of determinants, 
we  have:
\begin{equation}
\begin{split}
\mathcal{R}_{k}&=\underbrace{\text{B}\hspace{0.05cm}\text{log}_{2}\hspace{0.05cm}\begin{vmatrix}\sigma_{z}^{2}\textbf{L}_{k}^{H}\textbf{L}_{k}+\sum\limits_{l}^{K}\sum\limits_{m}\sum\limits_{m'}\sqrt{\eta_{m,l}\eta_{m',l}}\textbf{B}_{k,l,m}\textbf{B}_{k,l,m'}^{H}
	\end{vmatrix}}_{g_{1}(\boldsymbol{\eta})} \\
&-\underbrace{\text{B}\text{log}_{2}\hspace{0.05cm}\begin{vmatrix}\sigma_{z}^{2}\textbf{L}_{k}^{H}\textbf{L}_{k}+\sum\limits_{l\neq k}^{K}\sum\limits_{m}\sum\limits_{m'}\sqrt{\eta_{m,l}\eta_{m',l}}\textbf{B}_{k,l,m}\textbf{B}_{k,l,m'}^{H}\end{vmatrix}}_{g_{2}(\boldsymbol{\eta})} \; ,
\end{split}
\label{eq:Rate_final}
\end{equation}
where the vector $\boldsymbol{\eta}$ is a compact notation to denote the set of all the downlink transmit powers $\eta_{m, k}$, for all the values of $k$ and $m$.
All previous formulas concern the UC approach but they can be extended to the CF approach with ordinary efforts.

\subsection{Uplink data transmission}
The third phase of the communication protocol amounts to uplink data transmission. 
We  denote by $\textbf{x}_{k}^{UL}(n)$ the $P$-dimensional data vector to be transmitted by the $k$-th MS in the \emph{n}-th sample time;
the corresponding signal received at the \emph{m}-th AP is expressed as: 
\begin{equation}
\label{eq:y_m_UL}
\textbf{y}_{m}(n)=\sum\limits_{k=1}^{K}\sqrt{\widetilde{\eta}_{k}}\textbf{H}_{k,m}\textbf{L}_{k}\textbf{x}_{k}^{UL}(n)+\textbf{w}_m(n)
\end{equation}
where 
$\widetilde{\eta}_{k}=\dfrac{P_{t,k}^{UL}}{\mbox{tr}(\textbf{L}_{k}^{H}\textbf{L}_{k})}$,
and $P_{t,k}^{UL}$ is the uplink transmitted power by the $k$-th MS.

\subsubsection{CF approach} in this case, each AP forms the following statistic, $\forall k$: 
\begin{equation}
\label{eq:y_mk_tilde_UL}
\tilde{\textbf{y}}_{m,k}(n)=\textbf{Q}_{k,m}^H\textbf{y}_{m}(n) \; ,
\end{equation}
wherein now the ZF beamformer $\textbf{Q}_{k,m}$ previously defined is used as a post-coder.
Then each AP sends to the CPU the vectors $\tilde{\textbf{y}}_{m,k}(n)$ via the backhaul link, and the CPU forms the following soft estimate of the data vectors transmitted by th $k$-th MS:
\begin{equation}
\label{eq:x_k_hat_UL_Perfect CSI}
\hat{\textbf{x}}_{k}^{UL}(n)=\sum\limits_{m=1}^{M}\tilde{\textbf{y}}_{m,k}(n), \;\;\; k=1,\dots, K. 
\end{equation}
Plugging \eqref{eq:y_m_UL} and \eqref{eq:y_mk_tilde_UL} into \eqref{eq:x_k_hat_UL_Perfect CSI} it can be checked that the signal contributions associated to the data-symbol  $\textbf{x}_{k}^{UL}(n)$ are coherently summed.
\subsubsection{UC approach} in this case the signal transmitted by the \emph{k}-th MS is decoded only by the APs belonging to the set $\mathcal{M}(k)$. Accordingly, the CPU performs the following soft estimate: 
\begin{equation}
\label{x_k_ul_UC}
\hat{\textbf{x}}_{k}^{UL,UC}=\sum\limits_{m\in\mathcal{M}(k)}\tilde{\textbf{y}}_{m,k}(n), \;\;\; k=1,\dots,K,
\end{equation}
Substituting \eqref{eq:y_m_UL} and \eqref{eq:y_mk_tilde_UL} into \eqref{x_k_ul_UC}, we get: 
\begin{equation}
\begin{split}
\hat{\textbf{x}}_{k}^{UL,UC}&=\sum_{m \in {\cal M}(k)}\textbf{Q}_{k,m}^{H}\sqrt{\tilde{\eta}_{k}}\textbf{H}_{k,m}\textbf{L}_{k}\textbf{x}_{k}^{UL}(n)+\\
&+\sum\limits_{j=1, j\neq{k}}^{K}\sqrt{\tilde{\eta}_{j}}\sum_{m \in {\cal M}(k)}\textbf{Q}_{k,m}^{H}\textbf{H}_{j,m}\textbf{L}_{j}\textbf{x}_{j}^{UL}(n)+\\&+
\sum_{m \in {\cal M}(k)} \textbf{Q}_{k,m}^{H} \textbf{w}_{m}(n) \; .
\end{split}
\end{equation}
\medskip
Given the above equation, the achievable rate for the $k$-th MS can be easily shown to be expressed in the UC case as
\begin{equation}
\widetilde{\cal R}_k \!=\!B \log_2 \!\left| \mathbf{I}+ \widetilde{\eta}_k \mathbf{R}_k^{-1}
\!\!\left(\sum_{m \in {\cal M}(k)}\mathbf{B}_{k,k,m}^H\!\right)\!\!\! \left(\sum_{m \in {\cal M}(k)}\mathbf{B}_{k,k,m}\! \right)\!\!\right| \, ,
\end{equation}
wherein the interference covariance matrix now is written as
\begin{equation}
\begin{array}{lll}
\mathbf{R}_k=& \ds \sum_{j \neq k}\widetilde{\eta}_j\!\!
\left(\sum_{m \in {\cal M}(j)}\mathbf{B}_{j,k,m}^H\right)\!\!\!
\left(\sum_{m \in {\cal M}(j)}\mathbf{B}_{j,k,m}\right)  \\ \\
& +\sigma^2_w \mbox{card}({\cal M}(k) ) \; ,
\end{array}\end{equation}
with $\mbox{card}(\cdot)$ denoting cardinality. The above expression can be used also for the CF case by letting ${\cal M}(k)=\{1, 2, \ldots, M\}$, for all $k=1, \ldots, K$.

\section{Global Energy Efficiency Maximization}\label{Section:GEE}
In this section we address the problem of power control for energy efficiency maximization considering both the downlink and the uplink channel of the described cellular network. The downlink case is treated in the coming Section \ref{Sec:DownlinkGEE}, while the uplink scenario is dealt with in Section \ref{Sec:UplinkGEE}.

\subsection{Downlink power control}\label{Sec:DownlinkGEE}
Mathematically the problem is formulated as the optimization program:
\begin{subequations}\label{Prob:MaxGEE}
	\begin{align}
	&\max_{\boldsymbol{\eta}}\dfrac{\sum\limits_{k=1}^{K}\mathcal{R}_{k}({\boldsymbol{\eta}})}{ \sum\limits_{m=1}^{M}\left[\sum\limits_{k \in \mathcal{K}(m)} \delta\eta_{m,k} + P_{c,m}\right]}\label{Prob:MaxGEEa} \\
	&\text{s.t.}\sum\limits_{k \in \mathcal{K}(m)} \eta_{m,k} \leq P_{max,m}, \forall m=1,\dots,M \\
	&\quad\;\;  \eta_{m,k} \geq 0, \forall m=1,\dots,M, k=1,\dots,K 
	\end{align}
	\label{eq:max_GEE}
\end{subequations}
where $P_{c,m}>0$ is the circuit power consumed at AP $m$, $\delta\geq1$ is the inverse of the transmit amplifier efficiency, and $P_{max,m}$ is the maximum transmit power from AP $m$; $\boldsymbol{\eta}$ is a $KM\times 1$ vector containing all the transmit power of all AP.

Problem \eqref{Prob:MaxGEE} is challenging due to its fractional objective, which has a non-concave numerator. This prevents the direct use of standard fractional programming methods such as Dinkelbach's algorithm to solve \eqref{Prob:MaxGEE} with affordable complexity. In addition, another issue is represented by the large number of optimization variables, i.e. $KM$. To counter both issues, we resort to the successive lower-bound maximization method, introduced in \cite{Razaviyayn2013}\footnote{In \cite{Razaviyayn2013} the method is termed successive upper-bound minimization since there the focus is on minimization problems.} (it has already been used in other publications, \cite{sardellitti2014joint} and \cite{nguyen2013precoding}), and whose details are summarized in the Appendix. In brief, this method tackles \eqref{Prob:MaxGEE} by alternatively optimizing the transmit powers of each AP, while keeping the transmit powers of the other APs fixed. However, as it will be shown, even with respect to the transmit powers of a single AP, the global energy efficiency problem remains challenging and for this reason, each subproblem is tackled by means of sequential optimization, as described next.

Consider thus problem (\ref{eq:max_GEE}) and define the variable blocks 
$\boldsymbol{\eta}_m=\left\{ \eta_{m,k}\right\}_{k=1, \ldots, K}$, for the CF case, and 
$\boldsymbol{\eta}_m=\left\{ \eta_{m,k}\right\}_{k \in {\cal K}(m)}$, for the UC case; 
the vector $\boldsymbol{\eta}_m$ is $K$-dimensional in the CF case, while in the UC approach it will contain $N$ entries, i.e. the powers to be used to communicate with the MSs in the set ${\cal K}(m)$. The global energy efficiency maximization problem with respect to the $m$-th variable block refers to the maximization, with respect to $\boldsymbol{\eta}_m$ of the quantity
\begin{equation}
\dfrac{\sum\limits_{k=1}^{K}\mathcal{R}_{k}({\boldsymbol{\eta}_{m}},{\boldsymbol{\eta}_{-m}})}{ \sum\limits_{m=1}^{M}\left[\sum\limits_{k \in \mathcal{K}(m)} \delta \eta_{m,k} + P_{c,m}\right]}\; ,
\label{eq:33}
\end{equation}
with $\boldsymbol{\eta}_{-m}$ representing the set of all the APs' transmit powers except the ones in $\boldsymbol{\eta}_{m}$.
It can be easily seen that the numerator of the global energy efficiency is not concave even with respect to only the variable block $\boldsymbol{\eta}_m$.
In particular, \eqref{eq:33} has a non-concave numerator, since $\mathcal{R}_{k}(\boldsymbol{\eta})$, which depends on all transmit powers of the base stations, is the difference of two logarithmic functions with the powers $\boldsymbol{\eta_{m}}$ of AP $m$ appearing in both logarithmic terms, and the difference of two logarithms is in general not concave.
 This means that \eqref{eq:max_GEE} can not be solved with affordable complexity by either standard convex optimization nor fractional programming tools. Therefore, plain alternating maximization is not suitable, which motivates the use of the successive lower-bound maximization method. To this end, let us observe that each summand at the numerator of the global energy efficiency can be expressed as the difference between the functions $g_{1}$ and $g_{2}$ defined in \eqref{eq:Rate_final}. At this point we offer the following result.
\begin{lemma}\label{Lem:ConSqrt}
	Both functions $g_{1}$ and $g_{2}$ in \eqref{eq:Rate_final} are concave in $\boldsymbol{\eta}_{m}$ 
\end{lemma}
\begin{IEEEproof}
As a first step, let us show that the function $f: (x,y)\in\{\mathbb{R}_{0}^{+}\times \mathbb{R}_{0}^{+}\}\to \sqrt{xy}\in\mathbb{R}_{0}^{+}$ is jointly concave in $(x,y)$. To this end, the Hessian of $f$ can be written as
\begin{equation}
{\cal H}=\frac{1}{4}\left(\begin{array}{cc}-x^{-3/2}y^{1/2} & x^{-1/2}y^{-1/2} \\x^{-1/2}y^{-1/2} & -x^{1/2}y^{-3/2}\end{array}\right)\;,
\end{equation}
which can be seen to be negative-semidefinite, since its determinant is zero and the elements on the diagonal are non-positive. 

Next, we observe that both $g_{1}$ and $g_{2}$ in \eqref{eq:Rate_final} are obtained by composing the function $f$ with the $\text{log}_2 |(\cdot)|$ function of positive-definite argument. Then, the result follows recalling that the function $\text{log}_2 |(\cdot)|$ is matrix-concave and matrix-increasing over the set of positive-semidefinite matrices \cite[Section 3]{boyd2004convex}. 
\end{IEEEproof}
Based on Lemma \ref{Lem:ConSqrt}, we argue that (\ref{eq:Rate_final}) is the difference between two concave function. Thus, recalling that any concave function is upper-bounded by its first-order Taylor expansion around any given point $\boldsymbol{\eta}_{m,0}$, a concave lower-bound of $\mathcal{R}_{k}$ can be obtained as: 
\begin{equation}
\begin{split}
\mathcal{R}_{k}&=g_{1}(\boldsymbol{\eta}_{m})-g_{2}(\boldsymbol{\eta}_{m})\\
& \geq g_{1}(\boldsymbol{\eta}_{m})-g_{2}(\boldsymbol{\eta}_{m,0})-\nabla_{\boldsymbol{\eta}_{m}}^{T}g_{2}\mid_{\boldsymbol{\eta}_{m,0}}(\boldsymbol{\eta}-\boldsymbol{\eta}_{m,0})\\
& =\overline{\mathcal{R}}_k(\boldsymbol{\eta}-\boldsymbol{\eta}_{m,0}).
\end{split}
\label{eq:taylor_exp_rate}
\end{equation}  
Then, a suitable approximate problem for the implementation of the sequential optimization method is  
\begin{subequations}
	\begin{align}
	&\max_{\boldsymbol{\eta}_{m}}\dfrac{\sum\limits_{k=1}^{K}\overline{\mathcal{R}}_{k}({\boldsymbol{\eta}_{m}},{\boldsymbol{\eta}_{m,0}},{\boldsymbol{\eta}_{-m}})}{\delta \sum\limits_{m=1}^{M}\sum\limits_{k \in \mathcal{K}(m)} \eta_{m,k} + P_c} \\
	&\text{s.t.}\sum\limits_{k \in \mathcal{K}(m)} \eta_{m,k} \leq P_{max,m}\\
	& \quad\;\; \eta_{m,k} \geq 0, \; k=1,\dots,K 
	\end{align}
	\label{eq:max_GEE3}
\end{subequations}
It can be seen that Problem \eqref{eq:max_GEE3} fulfills the three properties \textbf{P1}, \textbf{P2} and \textbf{P3} of the sequential method detailed in the Appendix\footnote{The first-order Taylor expansion and its derivative coincide with the function and with its derivative when evaluated at the center of the expansion}. 
Moreover, the numerator of the objective is now concave, which enables the use of fractional programming tools to globally solve \eqref{eq:max_GEE3}, such as the popular Dinkelbach's algorithm \cite{ZapNow15}. The overall power control procedure is formulated as in Algorithm \ref{algorithm:dinkelback} and, based on the properties of the successive lower-bound maximization method reviewed in the Appendix, we can state the following result.
\begin{proposition}
Algorithm \ref{algorithm:dinkelback} monotonically improves the global energy efficiency value after each iteration and converges to a first-order optimal point of the original global energy efficiency maximization problem in \eqref{eq:max_GEE}.
\end{proposition}

\begin{algorithm}[t]
	\caption{Sequential algorithm for GEE maximization}
	\label{algorithm:dinkelback}
	\begin{algorithmic} [1]
		\STATE set i=0;
		\STATE choose any feasible $\boldsymbol{\eta}_{1},\dots,\boldsymbol{\eta}_{M}$;
		\REPEAT
		\FOR {$m=1,\dots,M$}
		\FOR {$k=1,\dots,K$}
		\STATE choose any feasible $\boldsymbol{\eta}_{m,0}$;
		\STATE Solve (\ref{eq:max_GEE3}) by Dinkelbach's algorithm and call $\boldsymbol{\eta}_{m}^{*}$ the solution;
		\STATE update $\boldsymbol{\eta}_{m,0}$: $\boldsymbol{\eta}_{m,0}=\boldsymbol{\eta}_{m}^{*}$
		\ENDFOR
		\ENDFOR
		\UNTIL{convergence}
	\end{algorithmic}
\end{algorithm}

\subsubsection{An alternative definition for the GEE on the downlink}
The definition of the global energy efficiency reported in \eqref{eq:max_GEE} considers a circuit power consumption that does not depend on the transmit power used by each base station. However, a base station that does not transmit will consume a lower circuit power, since it will switch to idle mode. We stress that such a circumstance can arise when maximizing the energy efficiency, since the marginal increase to the system achievable rate granted by their activation is overweighted by the increased network power consumption.
To account for this circumstance, the terms $P_{c,m}$ can be further detailed to depend on the actual used transmit power, namely defining for all $m=1,\ldots,M$,
\begin{equation}
P_{c,m}=\widetilde{P}_{c,m}(\mathbbm{1}[P_{T}(m)>0]+0.5(1-\mathbbm{1}[P_{T}(m)>0]))
\label{eq:Pcm}
\end{equation}
where 
$P_T(m)=\sum_k \eta_{m,k}$ is the power radiated by the $m$-th AP and 
$\mathbbm{1}[P_{T}(m)>0]$ is the indicator function of the set $[P_{T}(m)>0]$, being $1$ when $P_{T}(m)>0$ and 0 otherwise.  According to \eqref{eq:Pcm}, we assume that the circuit power consumption, equal to $\widetilde{P}_{c,m}$ when the AP is active, is halved when it does not radiate any power. 

While this more sophisticated power consumption model accounts for the lower circuit power consumption of idle access points, it leads to a non-differentiable global energy efficiency function, due to the presence of the indicator function $\text{I}$ in \eqref{eq:Pcm}. In this case Algorithm \ref{algorithm:dinkelback} is still guaranteed to monotonically increase the global energy efficiency value after each iteration, but no first-order optimality can be guaranteed upon convergence, owing to the non-differentiability of \eqref{eq:Pcm}.  Nevertheless, we should remark that it is also possible to approximate the indicator function in \eqref{eq:Pcm} by a smooth function, such as a sigmoid, thus recovering the first-order optimality property of the algorithm.

\subsection{Uplink power control}\label{Sec:UplinkGEE}
With regard to the uplink, 
the problem of global energy efficiency maximization is formulated as the optimization program:
\begin{subequations}\label{Prob:MaxGEE_up}
	\begin{align}
	&\max_{\boldsymbol{\eta}}\dfrac{\sum\limits_{k=1}^{K}\widetilde{\mathcal{R}}_{k}(\widetilde{\eta}_1, \ldots, \widetilde{\eta}_K)}{\left[\sum_{k=1}^K \delta
	\widetilde{\eta}_k + \widetilde{P}_{c,k}\right] }\label{Prob:MaxGEEa_up} \\
	&\text{s.t.} \quad \widetilde{\eta}_{k}{\mbox{tr}(\mathbf{L}_k^H 
	\mathbf{L}_k)} \leq P_{T,max}, \forall k=1,\dots,K \\
	&\quad\;\;  \, \, \, \widetilde{\eta}_{k} \geq 0, \forall k=1,\dots,K \, ,
	\end{align}
	\label{eq:max_GEE_up}
\end{subequations}
where $ P_{T,max}$ denotes the maximum transmit power for the MSs\footnote{For the sake of simplicity this power is assumed to be the same for all the MSs; this assumption however can be easily relaxed.}. 
It is easy to realize that the optimization problem \eqref{eq:max_GEE_up} has the same structure as \eqref{eq:max_GEE}, thus implying that it can be solved by using the same procedure that has been developed for the downlink. It is also worth noting that while in the downlink the number of variables to be optimized is  $MK$ in the CF case and $MN$ in the UC case, for the uplink only $K$ transmit powers are to be optimized, thus implying that the power optimization is much less computationally intensive on the uplink than on the downlink.

\section{Sum-rate maximization}
\label{Section:rate_maximization}
While the main focus of this work is on the maximization of the global energy efficiency function, it should be emphasized that the same approach can be adopted also for sum-rate maximization, which can be tackled as a special case of global energy efficiency maximization. Nonetheless, as detailed in the following, there are instances, when perfect CSI is available, in which the sum-rate maximization algorithm reduces to a standard convex optimization problem. 
Specifically, the global energy efficiency function reduces to the sum-rate function upon plugging $\delta=0$ and $P_{c}=1$ in \eqref{Prob:MaxGEEa}. Thus, in principle, sum-rate maximization can be performed by applying the simplified version of Algorithm \ref{algorithm:dinkelback} wherein standard convex optimization routines can be used to solve Problem (\ref{eq:max_GEE3}) when $\mu=0$ and $P_{c}=1$ in each iteration\footnote{Note that when $\mu=0$ and $P_{c}=1$, (\ref{eq:max_GEE3}) is no-longer a fractional program.}. 


Nevertheless, sum-rate maximization has one peculiarity that should be highlighted with respect to the global energy efficiency maximization scenario. With reference to the case in which perfect CSI is available, it is possible to consider the ZF precoder 
\begin{equation}\label{eq:cov_matix_ZF}
\textbf{R}_k=\sigma_{z}\textbf{L}_{k}^{H}\textbf{L}_{k}=\frac{N_{MS}}{P}\sigma_{z}\textbf{I}\;,
\end{equation}
thus removing multi-user interference. As a result, the sum-rate function simplifies to: 
\begin{equation}
\label{eq:sum_rate_ZF}
\begin{split}
\sum\limits_{k=1}^{K}\mathcal{R}_{k}&=\sum\limits_{k=1}^{K}\text{B}\hspace{0.05cm}\text{log}_{2}\hspace{0.05cm}\Big|
\textbf{I}+\left(\frac{N_{MS}}{P}\sigma_{z}^{2}\textbf{I}\right)^{-1} \\& \sum\limits_{m,m'\in\mathcal{M}(k)}\sqrt{\eta_{m,k}\eta_{m',k}}\textbf{C}_{k,m,m'}\Big|
\end{split}
\end{equation}
where $\textbf{C}_{k,m,m'}$ is: 
\begin{equation}
\label{eq:C_kmm}
\textbf{C}_{k,m,m'}=\textbf{L}_{k}^{H}\textbf{H}_{k,m}^{H}\textbf{Q}_{k,m}\textbf{Q}_{k,m'}^{H}\textbf{H}_{k,m'}\textbf{L}_{k}\;,
\end{equation}
which is a concave function by virtue of Lemma \ref{Lem:ConSqrt}. In this case, since the sum-rate is already concave and can be globally maximized with polynomial complexity by using standard convex programming theory.


\begin{figure}[h]
	\centering
	\includegraphics[scale=0.7]{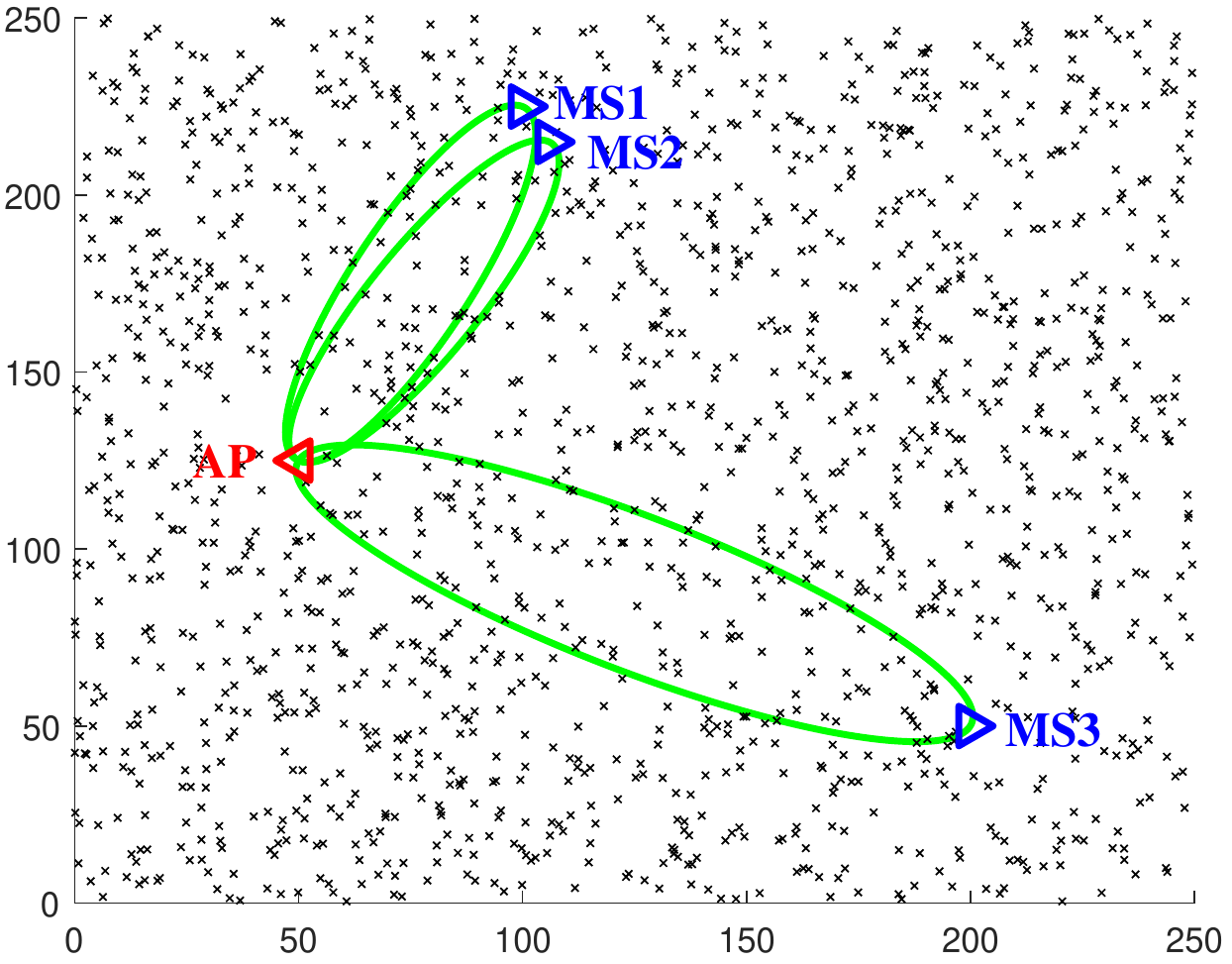}
	\caption{Detail about the channel generation procedure. For each AP-MS pair, only the scatterers falling into an ellipse built around the locations of the AP and of the MS are considered. This way, we can account for channel correlation when two receivers or two transmitters happen to be closely located. As an example, in the figure, the channels between the AP and MSs 1 and 2 will exhibit some sort of correlation, whereas the channel between the AP and the MS3 is statistically independent from the other channels.}
	\label{fig:ellipse}
\end{figure}

 \begin{figure}[t]
\centering
\includegraphics[scale=0.62]{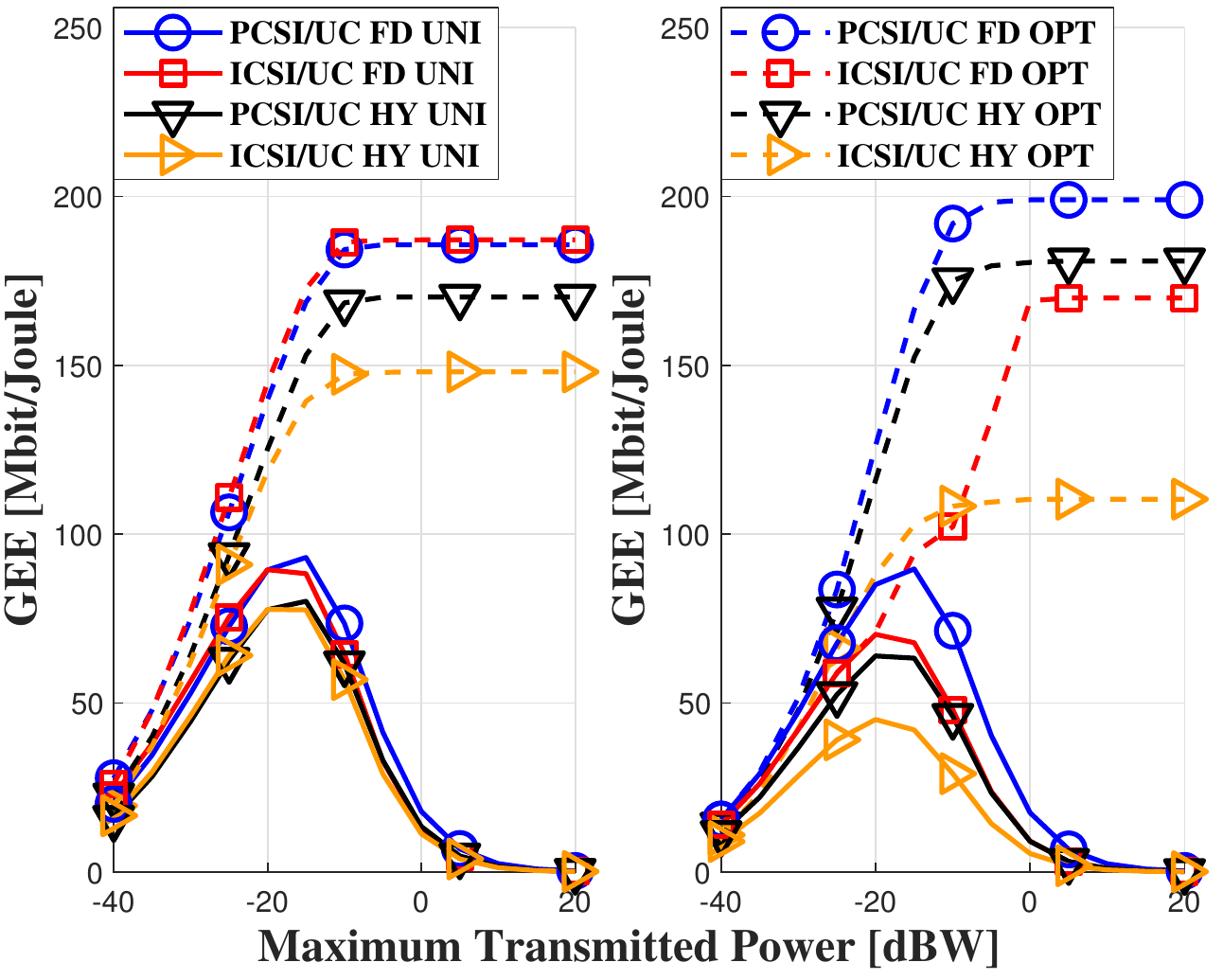}
\caption{Global Energy Efficiency with fully-digital (FD) and hybrid beamforming (HY) versus maximum transmit power. On the left we have the case N=1 and on the right the case N=3. System parameters: $M=80$, $K=6$, $N_{AP} \times N_{MS}=16 \times 8$, $P=1$, $\delta=1$, $P_{c}=1$ W.}
\label{fig:geeK6vspt}
\end{figure}

 \begin{figure}[t]
	\centering
	\includegraphics[scale=0.62]{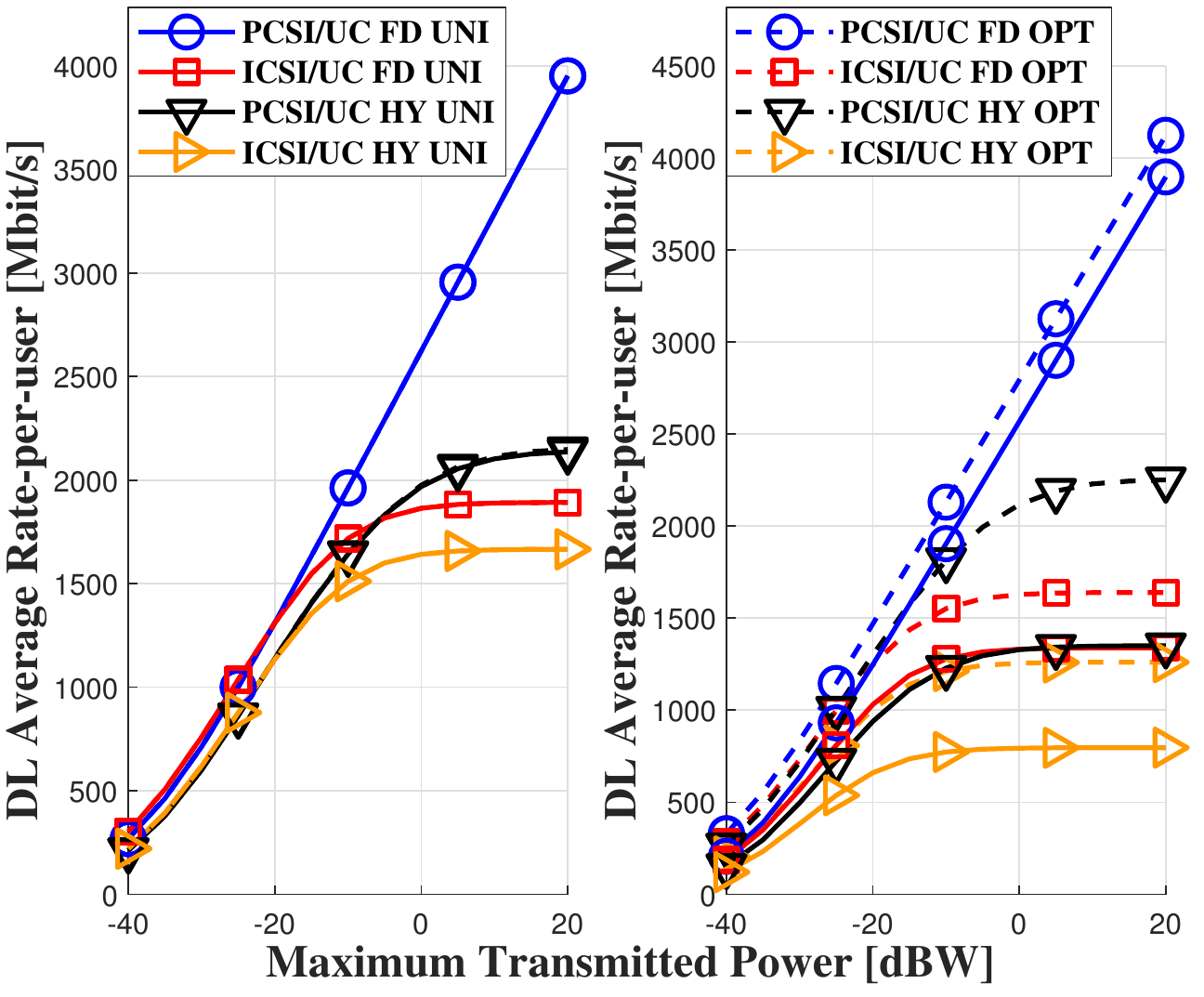}
	\caption{Average Achievable Rate-per-user with fully-digital (FD) and hybrid beamforming (HY) versus maximum transmit power. On the left we have the case N=1 and on the right the case N=3. System parameters: $M=80$, $K=6$, $N_{AP} \times N_{MS}=16 \times 8$, $P=1$.}
	\label{fig:rateK6vspt}
\end{figure}

\section{Performance analysis}
\label{Section:Numerical_results}

\subsection{Simulation setup}
We start by defining the used channel model for generating the matrices $\mathbf{H}_{k,m}$.
 According to the widely used clustered channel model for mmWave frequencies (see \cite{buzzidandreachannel_model} and references therein), $\textbf{H}_{k,m}$ can be expressed as 
\begin{equation}
\textbf{H}_{k,m}\!\!= \!\gamma\!\sum\limits_{i=1}^{N_{cl}}\!\sum\limits_{l=1}^{N_{ray}}\!\! \alpha_{i,l}\sqrt{L(r_{i,l})}\textbf{a}_{AP}(\theta_{i,l,k,m}^{AP})\textbf{a}_{MS}^{H}(\theta_{i,l,k,m}^{MS}) + \textbf{H}_{LOS},
\label{eq:HKM}
\end{equation}
where $N_{cl}$ is the number of clusters, $N_{ray}$ is the number of rays that we consider for each cluster, $\gamma$ is a normalization factor defined as $\sqrt{\dfrac{N_{AP}N_{MS}}{N_{cl}N_{ray}}}$, 
$\textbf{H}_{LOS}$ is the line-of-sight (LOS) component, $\alpha_{i,l}$ is the complex path gain distributed as $\mathcal{CN}(0,1)$, so that its amplitude is Rayleigh-distributed, $L(r_{i,l})$ is the attenuation related to the path $(i,l)$, $\textbf{a}_{AP}$ and $\textbf{a}_{MS}$ are the array responses at the $m$-th AP and at the $k$-th MS, respectively,  and they depend on the angles of arrival and departure, $\theta_{i,l,k,m}^{AP}$ and $\theta_{i,l,k,m}^{MS}$, relative to the 
$(i,l)$-th path of the channel between the $k$-th MS and the $m$-th AP.
The path-loss is defined as \cite{haneda20165g}
\begin{equation}
\label{attenuation_il}
L(r)=-20\log_{10}\biggl(\frac{4\pi}{\lambda}\biggr)-10n\log_{10}(r)-X_{\sigma},
\end{equation}
wherein $r$ is the distance between the transmitter and the receiver, $n$ is the path loss exponent, $X_\sigma$ is the shadow fading term in logarithmic units with zero mean and $\sigma^2$-variance and $f_0$  is a fixed frequency (see also table \ref{table:los}). The 
$\textbf{H}_{LOS}$ in (\ref{eq:HKM}) is written as\footnote{For the ease of notation we omit the subscript $k,m$.} 
\begin{equation}
\textbf{H}_{LOS}=I(d)\sqrt{N_{AP}N_{MS}}e^{j\eta}\sqrt{L(d)}\textbf{a}_{AP}(\theta_{LOS}^{AP})\textbf{a}_{MS}^{H}(\theta_{LOS}^{MS}).
\label{HLOS}
\end{equation}
In the above equation, $\eta\sim\mathcal{U}(0,2\pi)$, $I(d)$ is a 0-1 random variate indicating if a LOS link exists between the transmitter and the receiver, and $d$ is the transmitter-receiver distance, measured in meters. Denoting by $p$ the probability that $I_{LOS}(d)=1$, we have, for the  UMi (Urban Microcellular) scenarios \cite{5GBandsUpTo100}: 
\begin{equation}
p=\min\biggl(\frac{20}{d},1\biggr) (1-e^{-\frac{d}{39}})+e^{-\frac{d}{39}}. 
\end{equation} 
As for the number of scatterers and their positions, it should be said that, while usually for every AP-MS pair, a random and independently generated set of scatterers is considered to contribute to the channel matrix \eqref{eq:HKM}, in this paper, in order to model the possible channel correlation when the devices are closely spaced, we consider the same set of scatterers for the generation of all the channels. In particular, we assume that, in the considered area, there is a given number of random clusters, each one contributing with three rays\footnote{As specified in Section \ref{Section:Numerical_results}, a density of 0.4 cluster/sqm. will be considered.}.  Given these clusters, in order to generate the generic channel $\textbf{H}_{k,m}$ between the $k$-th MS and the $m$-th AP, we consider as active only those clusters falling in an ellipse built around the position of  the MS and the AP (see Fig. \ref{fig:ellipse}). In this way, on one hand we exclude far clusters from contributing to the channel, while, on the other hand, we are guaranteed that devices closely located will have correlated channels since they will be affected by similar sets of scatterers (see Fig. \ref{fig:ellipse} for a 
graphical illustration).

\medskip

\begin{table}[t] 	
	\caption{Parameters for Pathloss model}
	\center  	 
	\begin{tabular}{|c|c|}
		\hline 
		Scenario & Model Parameters \\ 
		\hline 
		UMi Street Canyon LOS & n=1.98, $\sigma$=3.1$\,$dB \\
		\hline 
		UMi Street Canyon NLOS & n=3.19, $\sigma$=8.2$\,$dB \\ 
		\hline 
		UMi Open Square LOS & n=2.89, $\sigma$=7.1$\,$dB \\ 
		\hline 
		UMi Open Square NLOS & n=1.73, $\sigma$=3.02$\,$dB  \\ 
		\hline 
	\end{tabular}  
	\label{table:los}
\end{table}
In the following simulation we have considered a carrier frequency of $f_0=73\,\textrm{GHz}$, a bandwidth of $B=200\,\textrm{MHz}$ and, with regard to the above channel model,  we have simulated the UMi Open Square scenario \cite{5GBandsUpTo100} of size $250 \times 250 \,\textrm{sqm}$. In order to generate the correlated channels, a total  of 25.000 randomly deployed clusters (corresponding to a cluster density of $0.4\,\textrm{cluster/sqm}$.)  has been generated. The additive white noise at the receiver has a power spectral density of $-174\,\textrm{dBm/Hz}$ and the receiver noise figure has been set to $F=6\,\textrm{dB}$. The simulated system has $M=80$ APs randomly deployed in the area to cover and equipped with an ULA with $N_{AP}=16$ antennas each. We consider both a scenario with $K=6$ users, that can be representative of a lightly-loaded network, and a scenario with $K=16$, that can be representative of a heavily-loaded scenario.  We plot results for the UC networking deployment with $N=1$ and $N=3$.
Each MS is equipped with and ULA with $N_{MS}=8$ antennas. Although the illustrated algorithms enable each MS  to transmit and receive multiple data-streams, a multiplexing order of $P=1$ is assumed here for the sake of simplicity. The presented results show the global energy efficiency [Mbit/Joule] as defined in \eqref{eq:max_GEE}, with the circuit power consumption expressed as in \eqref{eq:Pcm},   and the achievable rate per user [bit/s], defined as the sum-rate (i.e. the numerator of the global energy efficiency) divided by the number of MSs $K$.
For benchmarking purposes, we 
compare the performance achieved by the proposed power control rules (maximizing the global energy efficiency and the sum-rate) with 
that achieved by downlink uniform power control, which assumes that  
$\eta_{m,k}=P_T/K $
in the CF case and that 
\[
\eta_{m,k}=
\begin{cases}
\dfrac{{P_{t}}}{\text{card}(\mathcal{K}(m)) }, &k \in \mathcal{K}(m)\\
0, &k \notin \mathcal{K}(m)
\end{cases}
\]
for the UC approach, respectively.
The numerical values come from an average over 1000 independent channel scenarios as well as users and access points locations. 
The convergence criterion for the proposed sequential optimization algorithm is based on the computation of the \textit{relative tolerance}, i.e. the norm of the difference between the current optimized vector and the optimized vector available at the previous iteration, divided by the norm of the current optimized vector.  The successive convex approximation routine was directly implemented by the authors with a Matlab script, whereas the routine for maximizing the sum-rate in the case of concave rate was {\tt fmincon}.
All APs have been assumed to have the same maximum feasible transmit power $P_{max}$ on the downlink. The transmit amplifier efficiency of each transmitter has been assumed equal to one, i.e. $\delta=1$, while the hardware circuit power was modeled according to the model in \eqref{eq:Pcm} for each AP, with 
$\widetilde{P}_{c,m}=1\,\textrm{W}$, for all $m=1, \ldots, M$. For the uplink, we instead used $\widetilde{P}_{c,k}=0.3\,\textrm{W}$, for all $k=1, \ldots, K$.
In the figures, the dashed line represent the results obtained by means of optimization procedure, instead, the solid line (it has not been reported in the legend in order to avoid a redundancy), with the same color, represent the case of uniform power allocation.  

\subsection{Numerical results}

\begin{figure}[t]
	\centering
	\includegraphics[scale=0.62]{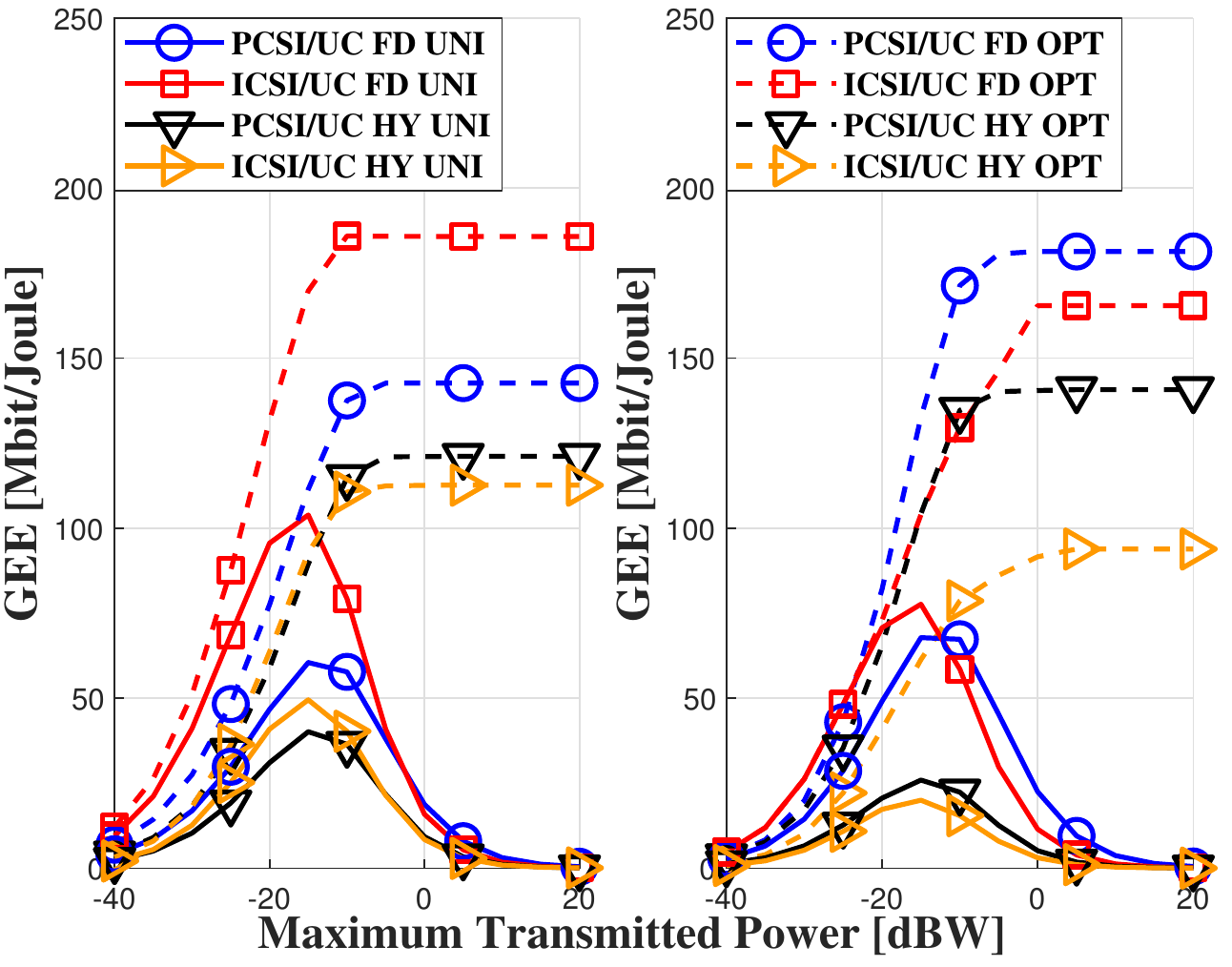}
	\caption{Global Energy Efficiency with fully-digital (FD) and hybrid beamforming (HY) versus maximum transmit power. On the left we have the case N=1 and on the right the case N=3. System parameters: $M=80$, $K=16$, $N_{AP} \times N_{MS}=16 \times 8$, $P=1$, $\delta=1$, $P_{c}=1$ W.}
	\label{fig:geeK16vspt}
\end{figure}

\begin{figure}[t]
	\centering
	\includegraphics[scale=0.62]{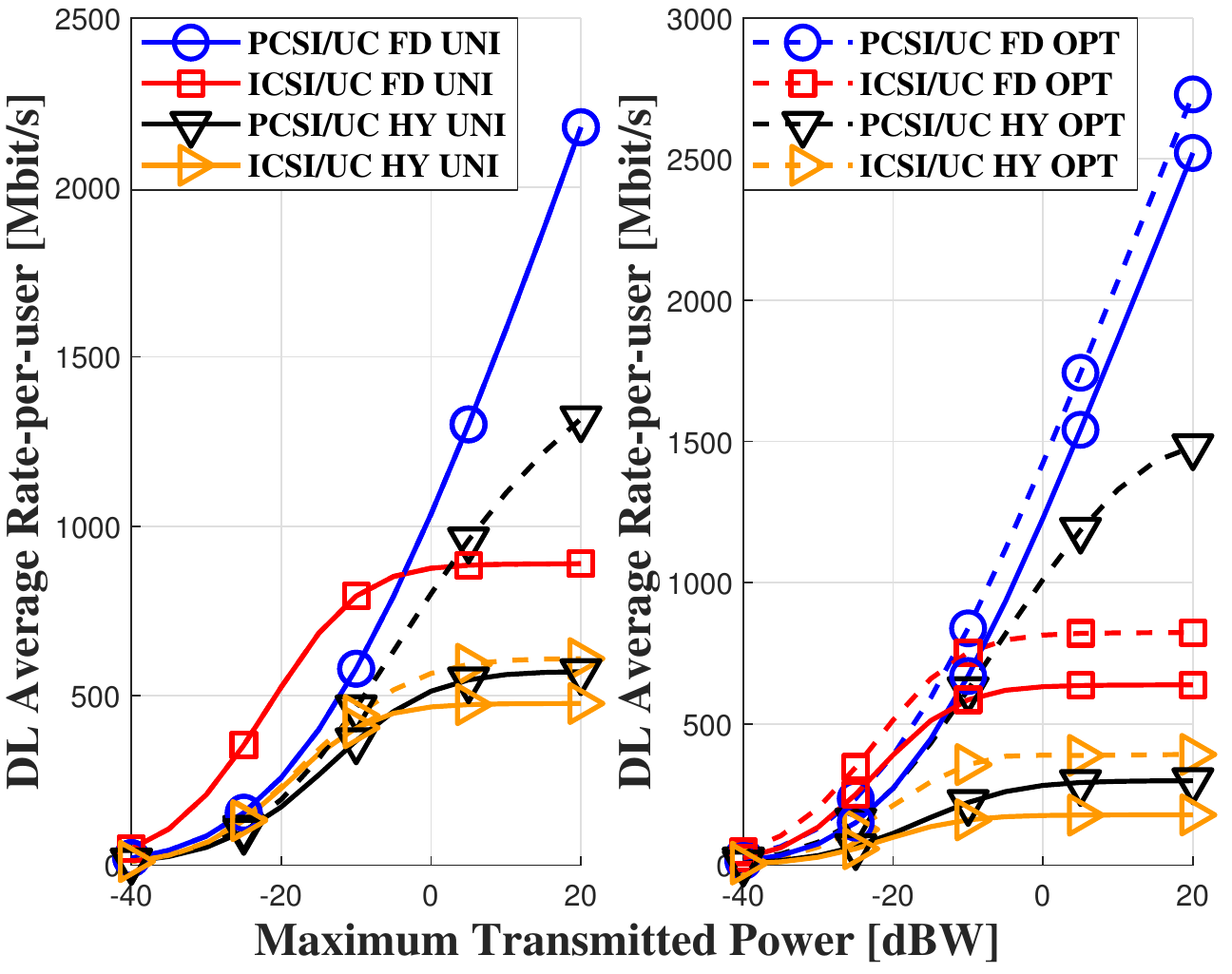}
	\caption{Average Achievable Rate-per-user with fully-digital (FD) and hybrid beamforming (HY) versus maximum transmit power. On the left we have the case N=1 and on the right the case N=3. System parameters: $M=80$, $K=16$, $N_{AP} \times N_{MS}=16 \times 8$, $P=1$.}
	\label{fig:rateK16vspt}
\end{figure}

Fig. \ref{fig:geeK6vspt} compares the global energy efficiency value versus $P_{max}$ achieved by the proposed 
global energy efficiency-maximizing power control scheme (labeled as OPT), considering the UC approach in the following scenarios:
\begin{itemize}
\item Perfect CSI and FD beamforming. 
\item Perfect CSI and HY beamforming, with $4$ RF chains used in the \emph{BCD-SD} HY beamforming algorithm. 
\item Imperfect CSI and FD beamforming, with pilot sequences of length $\tau_p=64$ and uplink transmit power of $100\,\textrm{mW}$.
\item Imperfect CSI and HY beamforming, with pilot sequences of length $\tau_p=64$ and uplink transmit power of $100\,\textrm{mW}$, with $4$ RF chains used in the \emph{BCD-SD} HY beamforming algorithm.
\end{itemize}
The plot on the left refers to the UC rule wth $N=1$, while on the right we have the case $N=3$. 
Fig. \ref{fig:rateK6vspt} considers the same setting as Fig. \ref{fig:geeK6vspt} but reports the downlink average rate per user achieved by the proposed rate-maximizing power control scheme (labeled as OPT). 
Figs. \ref{fig:geeK16vspt} and \ref{fig:rateK16vspt} report the same results as Figs. \ref{fig:geeK6vspt} and
\ref{fig:rateK6vspt}, respectively, with the difference that they refer to the highly-loaded scenario, i.e. with $K=16$ users. 
Inspecting the figures, several considerations can be made. First of all, we see that the proposed power optimization method provides better performance than the uniform power allocation scheme. This is not always true for the case with imperfect CSI (see plot on the left in Fig. \ref{fig:geeK16vspt}), as a consequence of the fact that, in the imperfect CSI case, the optimization step was performed using the estimated channels, while the plotted curve represents the true global energy efficiency, computed using the real channel coefficients.  In other words, as far as the imperfect CSI scenarios are concerned, the metric that is optimized is different from the metric that is plotted. Next, as expected, we notice that FD beamforming and the availability of perfect CSI lead to better performance with respect to the practical situation that imperfect CSI and HY beamforming is to be accounted for. For instance, focusing on the plot on the left in Fig \ref{fig:rateK6vspt} and considering a maximum transmit power of $0\,\textrm{dBW}$, the average rate-per-user drops from $2.65\, \textrm{Gbit/s}$ of the ideal case  to $1.65\, \textrm{Gbit/s}$, 
for the case in which both HY beamforming and ICSI are taken into account. Further, when comparing the lightly-loaded scenario with the highly-loaded one, we see that the system global energy efficiency does not dramatically change in the two situations, while, conversely, in a heavily-loaded system with HY beamforming and incomplete CSI the average rate-per-user at 0dBW of maximum transmit power is approximately equal to $500\, \textrm{Mbit/s}$, with about 60$\%$  loss with respect to the average rate-per-user attainable in a lightly loaded scenario.
A peculiar behavior that is observed in Figs. \ref{fig:rateK6vspt} and \ref{fig:rateK16vspt} is that in the case of perfect CSI with FD beamforming with $N=1$, the uniform and optimal power allocation schemes achieve the same performance. This is due to the fact that the rate is an increasing function of the total available power $P_{max}$ and so, when only one user needs to be served, uniform power allocation coincides with the rate-maximizing power allocation strategy. Instead, this behavior is not observed in  Figs. \ref{fig:geeK6vspt} and \ref{fig:geeK16vspt}, where a visible gap is present between the performance with uniform power allocation and with the optimized power allocation also when $N=1$. This is expected because, unlike the rate, the global energy efficiency is not monotonically increasing with $P_{max}$.

\begin{figure}[h]
	\centering
	\includegraphics[scale=0.62]{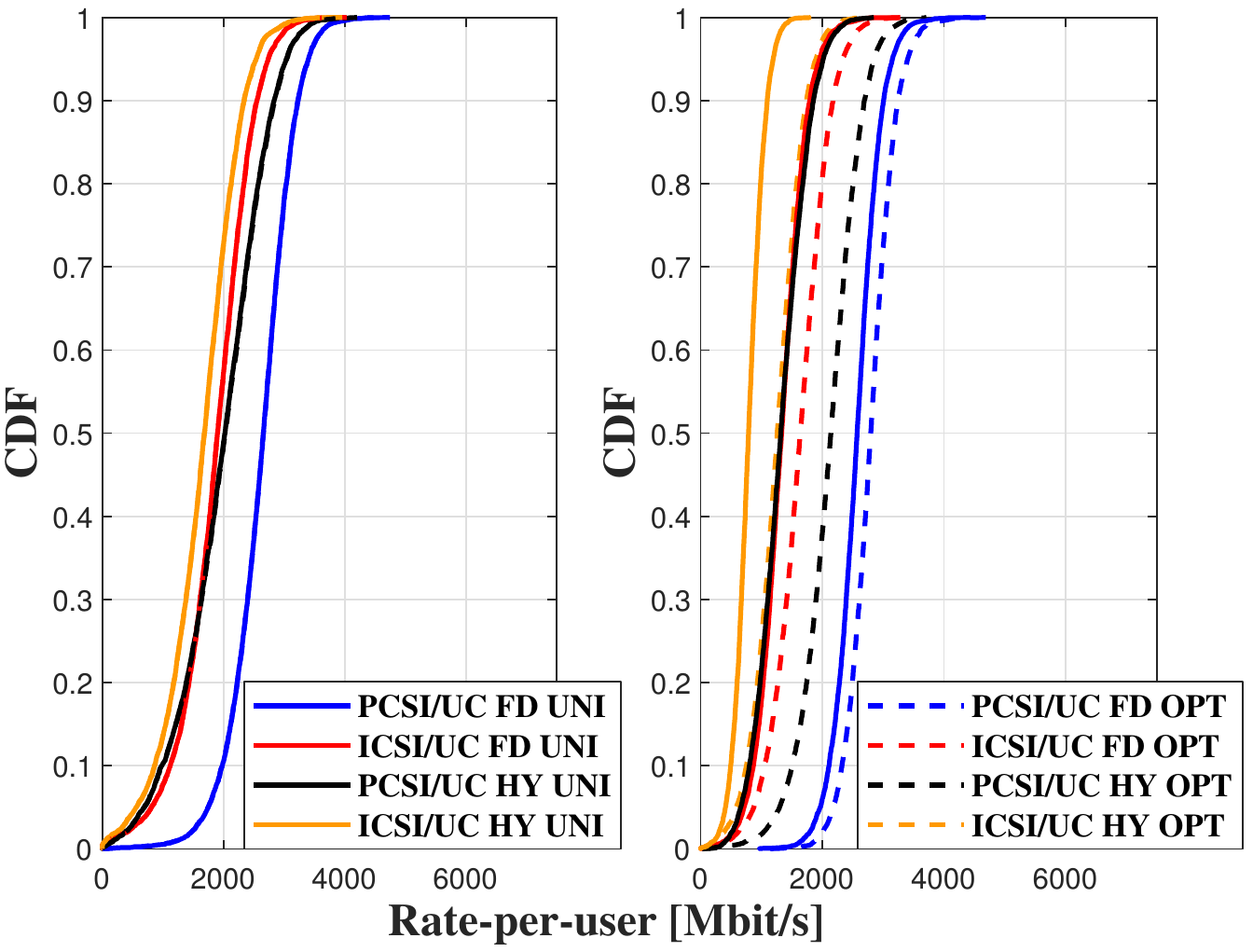}
	\caption{CDF of rate-per-user with fully-digital (FD) and hybrid beamforming (HY). On the left we have the case N=1 and on the right the case N=3. System parameters: $M=80$, $K=6$, $N_{AP} \times N_{MS}=16 \times 8$, $P=1$.}
	\label{fig:cdfrateK6uc}
\end{figure}

Next, Fig. \ref{fig:cdfrateK6uc} 
 shows the CDFs for the rate-per-user in case of FD and HY beamforming, respectively, considering $P_{max}=0\,\textrm{dBW}$ and the lightly loaded scenario. 
Again the two situations $N=1$ and $N=3$ are examined.  The curves confirm the findings previously commented. Additionally, it can be seen that for the practical case of incomplete CSI and HY beamforming the scenario with $N=1$ provides much better performance than the one with $N=3$ for the vast majority of the users, i.e. it appears to be convenient to have each AP serve just one user than 3 users, except than when considering the unlucky users that happen to be situated on the left tail of the rate-per-user distribution. This behavior can be explained by noticing that in the considered setting, where $M>>K$, the majority of the MSs has in its neighborhood several APs, and letting these APs dedicate their own resources to only MS results in overall increased performance. 
 
\begin{figure}[h]
	\centering
	\includegraphics[scale=0.62]{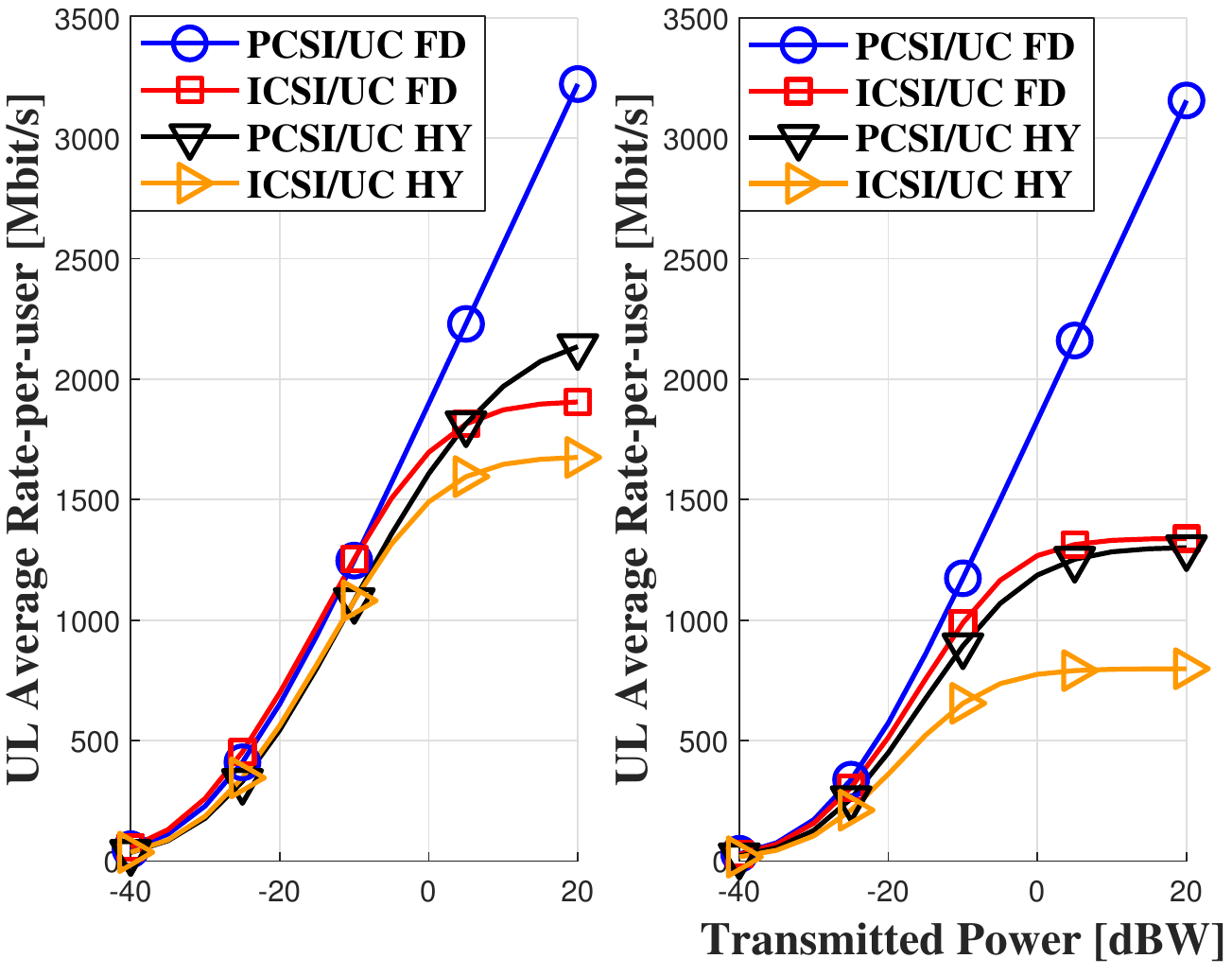}
	\caption{Uplink Average Achievable Rate-per-user with fully-digital (FD) and hybrid beamforming (HY) versus transmitted power. On the left we have the case N=1 and on the right the case N=3. System parameters: $M=80$, $K=6$, $N_{AP} \times N_{MS}=16 \times 8$, $P=1$.}
	\label{fig:avrateulk6uc}
\end{figure}

Finally, we turn our attention to the performance of the uplink channel. Fig. 
\ref{fig:avrateulk6uc} 
reports the average uplink rate-per-user versus the uplink transmit power. No power control is performed here. Again, the scenarios $N=1$ and $N=3$ are considered. From the figure, it is seen that also for the uplink the rate is a clearly increasing function of the transmit power only in the ideal case of perfect CSI and fully digital beamforming, while, in the other situations, a saturation effect is observed. Results also show that, when focusing on the practical scheme with incomplete CSI and HY beamforming the case $N=1$ provides better results than the case $N=3$. In particular, assuming an uplink transmit power of -10dBW a data-rate of about $1.2\,\textrm{Gbit/s}$ can be achieved for the case $N=1$, versus a data-rate of about $600\,\textrm{Mbit/s}$ for the case $N=3$. Again, this behavior can be explained by noticing that in a scenario with a dense AP deployment each MS is sorrounded by several APs, and it is better to let each AP to focus its own resourced on just one MS rather then letting the AP to share its resources among multiple MSs.

\section{Conclusion}
\label{Section:Conclusion}
This paper has considered a CF massive MIMO system operating at mmWave frequencies with a UC association between APs and MSs. Adopting a clustered channel model capable of taking into account the channel correlation for nearby devices, the paper has analyzed both the CF and UC approaches, by proposing a low-complexity power allocation rule aimed at global energy efficiency maximization. The presence of HY beamforming architectures at the APs has been considered, whereas, for the MSs, a simple channel-independent 0-1 beamforming structure has been considered. The obtained results have confirmed that the proposed resource allocation algorithms are effective at increasing the system energy efficiency and the system average rate-per-user,  as well as that the use of HY beamforming architectures introduces, as expected,  a considerable performance degradation. 
To the best of the authors' knowledge, this is the first paper to consider the CF and UC massive MIMO architectures at mmWave frequencies. This study can be generalized along several paths. First of all, a comparison with a massive MIMO deployment with co-located antennas should be carried out. Then, given the gap observed between the HY and the FD beamforming schemes, more sophisticated HY beamformers should be considered to improve the system performance. Finally, given the intrinsic macro-diversity guaranteed by the dense AP deployment, it would be of interest to investigate on the suitability of CF architectures to circumvent the blockage effects that cause link instability: in this sense, CF massive MIMO could be envisioned as an architecture capable  of making mmWave frequencies useful for ultra-reliable communications in indoor and small-sized outdoor environments.

\section*{Appendix: Successive lower-bound maximization}
In this appendix we briefly review the successive lower-bound optimization technique. 
The method is based on the  idea of merging the tools of alternating optimization \cite[Section 2.7]{BertsekasNonLinear} and sequential convex programming \cite{SeqCvxProg78}. Specifically, let us consider the generic optimization problem 
\begin{align}\label{Prob:GeneralProb}
\ds\max_{\boldsymbol{x}\in{\cal X}} f(\boldsymbol{x})\;,
\end{align}
with $f:\mathbb{R}^{n}\to \mathbb{R}$ a differentiable function, and ${\cal X}$ a compact set. Similarly to the alternating optimization method, the successive lower-bound maximization partitions the variable space into $J$ blocks, $\boldsymbol{x}=(\boldsymbol{x}_1,\ldots,\boldsymbol{x}_J)$, which are cyclically optimized one at a time, while the other blocks are kept fixed. Thus, Problem \eqref{Prob:GeneralProb} is decomposed into $M$ subproblems, wherein the generic subproblem is stated as
\begin{align}\label{Prob:SubProb}
\ds\max_{\boldsymbol{x}_{m}} f(\boldsymbol{x}_{m},\boldsymbol{x}_{-m})\;,
\end{align}
with $\boldsymbol{x}_{-m}$ cotaining all variable blocks except the $m$-th. If Problem \eqref{Prob:SubProb} is globally solved for each $m=1,\ldots,J$, then we have an instance of the alternating maximization method, which, as proved in \cite[Proposition 2.7.1]{BertsekasNonLinear}, monotonically improves the objective of \eqref{Prob:GeneralProb}, and yields a first-order optimal point if the solution of \eqref{Prob:SubProb} is unique for any $m$, and if ${\cal X}={\cal X}_1\times \ldots \times{\cal X}_J$, with $\boldsymbol{x}_{m}\in{\cal X}_{m}$ for all $m$. However, if globally solving \eqref{Prob:SubProb} is difficult (e.g. because \eqref{Prob:SubProb} is not a convex problem), then implementing the alternating maximization method proves difficult. In this case, the successive lower-bound maximization method proposes to find a (possibly suboptimal) solution of \eqref{Prob:SubProb}, by means of the sequential convex programming method. Besides leading to a computationally viable algorithm, this approach is proved to preserve the optimality properties of the alternating optimization method \cite{Razaviyayn2013}, despite the fact that a possible suboptimal solution of \eqref{Prob:SubProb} is determined. 

As for sequential optimization, its basic idea is to tackle a difficult maximization problem by a sequence of easier maximization problems. To elaborate, denote by $g_i(\boldsymbol{x}_{m})$ the $i$-th constraint of \eqref{Prob:SubProb},  for $i=1,\ldots,C$, and consider a sequence of approximate problems $\{{\cal P}_{\ell}\}_{\ell}$ with objectives $\{f_{\ell}\}_{\ell}$ and constraint functions $\{g_{i,\ell}\}_{i=1}^C$, such that the following three properties are fulfilled, for all $\ell$:
\begin{enumerate}
	\item[(\textbf{P1})] $f_{\ell}(\boldsymbol{x}_m)\leq f(\boldsymbol{x}_m)$, $g_{i,\ell}(\boldsymbol{x}_m)\leq g_{i,\ell}(\boldsymbol{x}_m)$, for all $i$ and $\boldsymbol{x}_m$;
	\item[(\textbf{P2})] $f_{\ell}(\boldsymbol{x}_m^{(\ell-1)})=f(\boldsymbol{x}_m^{(\ell-1)})$, $g_{i,\ell}(\boldsymbol{x}_m^{(\ell-1)})=g_{i}(\boldsymbol{x}_m^{(\ell-1)})$ with $\boldsymbol{x}_m^{(\ell-1)}$ the maximizer of $f_{\ell-1}$;
	\item[(\textbf{P3})] $\nabla f_{\ell}(\boldsymbol{x}_m^{(\ell-1)})=\nabla f(\boldsymbol{x}_m^{(\ell-1)})$, $\nabla g_{i,\ell}(\boldsymbol{x}_m^{(\ell-1)})=\nabla g_{i}(\boldsymbol{x}_m^{(\ell-1)})$.
\end{enumerate}
In \cite{SeqCvxProg78} (see also \cite{Beck2010,Razaviyayn2013}) it is shown that, subject to constraint qualifications, the sequence $\{f(\boldsymbol{x}_m^{(\ell)})\}_{\ell}$ of the solutions of the $\ell$-th Problem ${\cal P}_{\ell}$, is monotonically increasing and converges. Moreover, every convergent sequence $\{\boldsymbol{x}_m^{(\ell)}\}_{\ell}$ attains a first-order optimal point of the original Problem \eqref{Prob:SubProb}. Thus, sequential optimization guarantees at the same time to monotonically improve the objective function, and to fulfill the \gls{kkt} first-order optimality conditions of the original problem. Nevertheless, in order to be able to use the method, it is necessary to find lower bounds of the original objective function, which fulfill all three properties \textbf{P1}, \textbf{P2}, \textbf{P3}, while at the same time leading to an approximate problem that can be solved with affordable complexity.

\bibliographystyle{IEEEtran}
\bibliography{ref_TGCN}

\end{document}